\documentclass[useAMS,usenatbib]{mn2e}
\usepackage{graphicx}
\usepackage{rotating}
\usepackage{lscape}
\DeclareMathAlphabet{\mathpzc}{OT1}{pzc}{m}{it}

\def\lsim{\,\lower2truept\hbox{${<\atop\hbox{\raise4truept\hbox{$\sim$}}}$}\,}
\def\gsim{\,\lower2truept\hbox{${> \atop\hbox{\raise4truept\hbox{$\sim$}}}$}\,}

\title[Mid-/far-IR spectroscopic surveys]{Exploring the relationship between black hole accretion and star formation with blind mid-/far-infrared spectroscopic surveys}
\author[M. Bonato et al.]
{M. Bonato$^{1,2}$\thanks{matteo.bonato@oapd.inaf.it},
M. Negrello$^{2}$,
Z.-Y. Cai$^{3,4}$,
G. De Zotti$^{2,4}$,
A. Bressan$^{4}$,
A. Lapi$^{4,5}$,
\newauthor
F. Pozzi $^{6}$,
C. Gruppioni$^{7}$
and L. Danese$^{4}$ \\
$^{1}$Dipartimento di Fisica e Astronomia ``G.Galilei'', Universit\`a degli Studi di Padova, Vicolo Osservatorio 3, I-35122 Padova, Italy \\
$^{2}$INAF, Osservatorio Astronomico di Padova, Vicolo Osservatorio 5, I-35122 Padova, Italy \\
$^{3}$Center for Astrophysics, University of Science and Technology of China, Hefei, 230026, China \\
$^{4}$SISSA, Via Bonomea 265, I-34136 Trieste, Italy \\
$^{5}$Dipartimento di Fisica, Universit\`a ``Tor Vergata'', Via della Ricerca Scientifica 1, I-00133 Roma, Italy \\
$^{6}$Dipartimento di Fisica e Astronomia, Universit\`a di Bologna, via Ranzani 1, I-40127 Bologna, Italy \\
$^{7}$INAF, Osservatorio Astronomico di Bologna, Via Ranzani 1, I-40127 Bologna, Italy}
\date{Released 2014 Xxxxx XX}

\pagerange{\pageref{firstpage}--\pageref{lastpage}} \pubyear{2013}

\def\LaTeX{L\kern-.36em\raise.3ex\hbox{a}\kern-.15em
    T\kern-.1667em\lower.7ex\hbox{E}\kern-.125emX}
\def\simlt{\mathrel{\rlap{\lower 3pt\hbox{$\sim$}}\raise 2.0pt\hbox{$<$}}}
\def\simgt{\mathrel{\rlap{\lower 3pt\hbox{$\sim$}}\raise 2.0pt\hbox{$>$}}}

\begin{document}

\label{firstpage}

\maketitle

\begin{abstract}
We present new estimates of redshift-dependent luminosity functions of IR lines detectable by SPICA/SAFARI and excited both by star formation and by AGN activity. The new estimates improve over previous work by using updated evolutionary models and dealing in a self consistent way with emission of galaxies as a whole, including both the starburst and the AGN component. New relationships between line and AGN bolometric luminosity have been derived and those between line and IR luminosities of the starburst component have been updated. These ingredients were used to work out predictions for the source counts in 11 mid/far-IR emission lines partially or entirely excited by AGN activity. We find that the statistics of the emission line detection of galaxies as a whole is mainly determined by the star formation rate, because of the rarity of bright AGNs. We also find that the slope of the line integral number counts is flatter than 2 implying that the number of detections at fixed observing time increases more by extending the survey area than by going deeper. We thus propose a wide spectroscopic survey of 1\,hour integration per field-of-view over an area of 5\,deg$^{2}$ to detect  (at 5$\sigma$) $\sim$760 AGNs in [OIV]25.89$\mu$m $-$ the brightest AGN mid-infrared line $-$ out to $z\sim2$. Pointed observations of strongly lensed or hyper-luminous galaxies previously detected by large area surveys such as those by Herschel and by the SPT can provide key information on the galaxy-AGN co-evolution out to higher redshifts.
\end{abstract}


%%%%%%%%%%%%%%%%%%
\section{Introduction}\label{sect:intro}
%%%%%%%%%%%%%%%%%%

One of the main scientific goals of the SPace InfraRed telescope for Cosmology and Astrophysics (SPICA)\footnote{http://www.ir.isas.jaxa.jp/SPICA/SPICA\_HP/index-en.html} with its SpicA FAR infrared Instrument \citep[SAFARI;][]{Roelfsema12} is to provide insight into the interplay between galaxy evolution and the growth of active nuclei (AGNs) at their centers during the dust-enshrouded active star formation and mass accretion phases. The imaging spectrometer SAFARI is designed to fully exploit the extremely low far-infrared (far-IR) background environment provided by the SPICA observatory, whose telescope will be actively cooled at 6K. In each integration it will take complete 34--210\,$\mu$m spectra split in three bands ($\,34-60\,\mu$m, $\,60-110\,\mu$m, $110-210\,\mu$m), spatially resolving the full 2$^{\prime}\times$2$^{\prime}$ field of view (FoV). These bands are weakly if at all affected by dust extinction and thus allowing direct measurements of the intrinsic properties of the sources.

The mid-IR (MIR) lines that will be detected can come either from star forming regions or from nuclear activity or from both. Distinguishing the two contributions is not easy. However a key diagnostic relies on the fact that AGNs produce harder radiation and therefore excite metals to higher ionization states than starbursts. Thus lines with high-ionization potential are a powerful tool for identifying AGN activity \citep{Sturm02,Melendez2008,Satyapal2008,Satyapal2009,GouldingAlexander2009,Dudik2009,Weaver2010} that is easily missed by optical spectroscopic observations in the case of dusty objects, as galaxies with intense star formation are.

So far these studies have been limited to local objects. Substantial progress in this area will be made possible by spectroscopy with SAFARI that will exploit for the first time the rich suite of mid- and far-IR diagnostic lines to trace the star formation and the accretion onto the super-massive black holes up to high redshifts through both blind spectroscopic surveys and pointed observations.

An exploratory study of the SPICA/SAFARI potential in detecting galaxies and AGNs in the main IR lines in several redshift ranges has been presented by \citet{Spin12}. Their calculations were based on the phenomenological evolutionary models by \citet{Franc10}, \citet{Grupp11} and \citet{Valiante09}. The \citet{Franc10} and the \citet{Grupp11} models, like most classical backward evolution models, evolve the star forming galaxies and the AGNs independently. This class of models have been quite successful in reproducing far-IR to sub-millimeter source counts but are clearly unable to predict the composite line spectra of galaxies in which star formation and AGN activity co-exist.

\citet{Valiante09} have carried out a first attempt to deal in a coherent way with the cosmic evolution of IR emissions of both the starburst and the AGN components. This was done by first estimating the distribution of the ratios between the $6\,\mu$m AGN continuum luminosity, $\nu L_{6\mu\rm m}$, and the total IR luminosity, $L_{\rm IR}=L_{8-1000\mu\rm m}$, for a sample of local galaxies in five IR luminosity bins. The luminosity dependent distributions of $\nu L_{6\mu\rm m}/L_{\rm IR}$ were derived subtracting the estimated starburst contribution from the \textit{Spitzer} Infrared Spectrograph (IRS) mid-IR data for a complete sample of IRAS galaxies. A check on the evolution with redshift of the mean $\nu L_{6\mu\rm m}$ vs $L_{\rm IR}$ relation was made using IRS measurements for a sample of galaxies at $0.37 < z < 3.35$. The mean ratio $\nu L_{6\mu\rm m}/L_{\rm IR}$ was found to increase with luminosity as $\nu L_{6\mu\rm m}/L_{\rm IR}\propto L_{\rm IR}^{1.4\pm 0.6}$. The increase was assumed to stop at $L_{\rm IR,f}=10^{12.8}\,L_\odot$ for $z<0.5$ and at $L_{\rm IR,f}=10^{12}\,L_\odot$ for $z\ge 0.5$. Then an evolutionary law for $L_{\rm IR}$ was chosen and a value of $\nu L_{6\mu\rm m}/L_{\rm IR}$ drawn at random from the appropriate distribution was assigned to each source.

The strong increase of the average AGN luminosity fraction with increasing $L_{\rm IR}$ up to $L_{\rm IR,f}$ derived by \citet{Valiante09} is however not supported by many other analyses, although the relationship between these two quantities is still matter of debate with studies showing either strong or marginal correlations between AGN and starburst luminosities  \citep{Lutz08,Serjeant09,AlexanderHickox2012,DiamondStanicRieke2012,Rosario12,Mullaney12a,Mullaney12b,Chen13,ChenHickox2014,Hickox14}. The apparently contradictory results of different analyses may be understood taking into account on one side that the characteristic timescale of black hole accretion is very different from that of star formation and on the other side that any relation between star formation rate (SFR) and accretion rate must break down as the AGN luminosity is at the Eddington limit.

The different timescales imply that accretion and star formation are not necessarily on at the same time. Hence the relation among the two quantities may have a large dispersion when individual objects are considered, still being strong when averaging over the whole population of star forming galaxies \citep{Chen13}. A coherent scenario for the interpretation of the various pieces of evidence on the connection between star formation and AGN activity has been elaborated by \citet{Lapi2014}.

In this paper we present a new approach to the problem taking advantage of the physical model for the co-evolution of super-massive black holes and massive proto-spheroidal galaxies at high redshift ($z\ge 1.5$) worked out by \citet{Cai13} who presented the evolving luminosity functions of these objects as a whole (starburst plus AGN), taking into account in a self-consistent way the variation with galactic age of the global spectral energy distribution.

The physical model was complemented by a phenomenological model for lower $z$ AGNs and star forming galaxies, mostly late-type since massive spheroidal galaxies are observed to be in essentially passive evolution at $z \lsim 1$--1.5. In this case galaxies and AGNs were evolved separately. We combined the two components exploiting the average accretion rate as a function of the SFR derived by \citet{Chen13}, taking into account its dispersion. Bright optically selected QSOs, that do not obey such correlation, were taken into account adopting the best fit evolutionary model by \citet{Croom2009} up to $z=2$.

The plan of the paper is the following. In Section\,\ref{sect:evol} we present the adopted model for the evolution with cosmic time of the  IR (8-1000\,$\mu$m) luminosity function. In Section\,\ref{sect:line_vs_IR} we discuss the relations between line and continuum luminosity for the main mid/far-IR lines. In Section\,\ref{sect:LF} we work out our predictions for line luminosity functions, number counts and redshift distributions in the SPICA/SAFARI bands. In Section\,\ref{sect:comparison} we compare our results with previous estimates. In Section\,\ref{sect:survey} we discuss possible SPICA/SAFARI observation strategies, considering different integration times per FoV. Section\,\ref{sect:concl} contains a summary of our main conclusions.

We adopt a flat $\Lambda \rm CDM$ cosmology with matter density $\Omega_{\rm m} = 0.32$, dark energy density $\Omega_{\Lambda} = 0.68$ and Hubble constant $h=H_0/100\, \rm km\,s^{-1}\,Mpc^{-1} = 0.67$ \citep{PlanckCollaborationXVI2013}.

\begin{figure*}
\includegraphics[trim=2.2cm 2.2cm 2.7cm 2.7cm,clip=true,width=\textwidth]{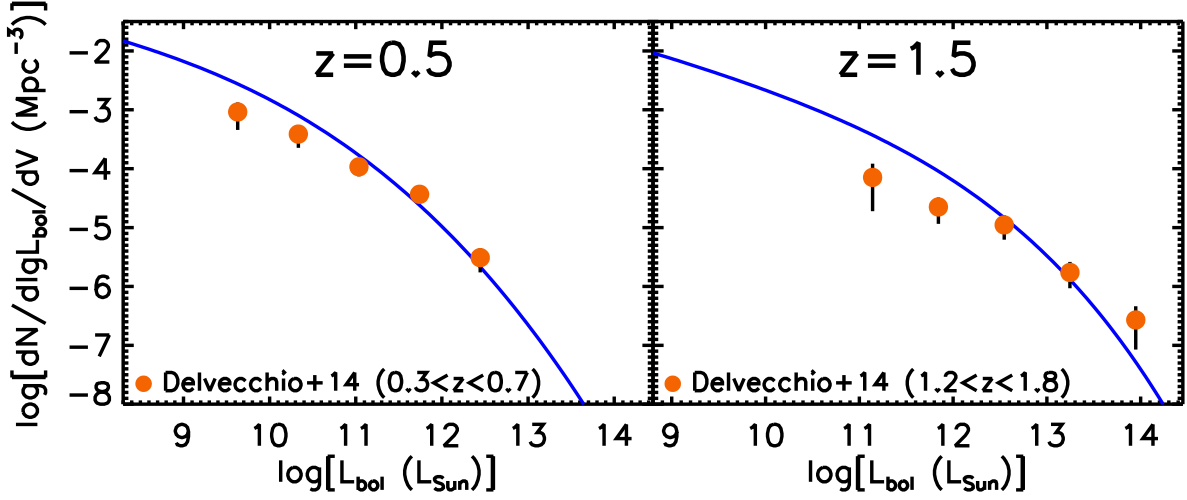}
\caption{Comparison of the AGN bolometric luminosity functions at $z=0.5$ and $z=1.5$ yielded by our approach (solid lines) with the observational estimates by \citet{Delvecchio14}. The dotted red curve in the right-hand panel shows the contribution of strongly lensed (amplification $\mu \ge 2$) sources; at $z=0.5$ such contribution is negligible. The vertical lines correspond to the minimum bolometric luminosity yielding a detection by SAFARI of the strongest AGN line (${\rm [OIV]}\,25.89\mu$m) with exposures of 1 and 10 hr/FoV (solid and dotted, respectively), for the mean line to bolometric luminosity ratios.}
 \label{fig:LF_bol}
\end{figure*}

%%%%%%%%%%%%%%%%%%%%%%%%%%%%%%%%%%%%%%%%%%%
\section{Evolution of the IR luminosity functions}\label{sect:evol}
%%%%%%%%%%%%%%%%%%%%%%%%%%%%%%%%%%%%%%%%%%%

As mentioned in Sect.~\ref{sect:intro} our reference model for the redshift-dependent IR luminosity functions is the one worked out by \citet{Cai13} based on a comprehensive ``hybrid'' approach  combining a physical model for the progenitors of early-type galaxies with a phenomenological one for late-types. The evolution of progenitors of early-type galaxies and of massive bulges of Sa's is described by an updated version of the physical model by \citet[][see also Lapi et al. 2006, 2011 and Mao et al. 2007]{Granato2004}. In the local universe these objects are composed of relatively old stellar populations with mass-weighted ages $\gsim 8$--9\,Gyr, corresponding to formation redshifts $z\gsim 1$--1.5, while the disc components of spirals and the irregular galaxies are characterized by significantly younger stellar populations \citep[cf.][their Fig.~10]{Bernardi2010}. Thus the progenitors of early-type galaxies, referred to as proto-spheroidal galaxies or ``proto-spheroids'', are the dominant star forming population at $z\gsim 1.5$, while IR galaxies at $z\lsim 1.5$ are mostly late-type galaxies.

According to \citet{Cai13} the redshift dependent luminosity functions of proto-spheroids as a whole, $\Phi(L_{\rm tot},z)$,  writes:
\begin{eqnarray}\label{eq:prob3}
&& \Phi(L_{\rm tot},z) = \int_{M{\rm min}}^{M_{\rm max}} dM_{\rm vir} \int_{z}^{z_{\rm max}} dz_{\rm vir} \bigg\vert \frac{dt_{\rm vir}}{dz_{\rm vir}} \bigg\vert \nonumber \\
&& \times \frac{dN_{ST}}{dt_{\rm vir}} (M_{\rm vir},z_{\rm vir}) \int_{-\infty}^{\log L_{\rm tot}} \frac{dx}{2\pi\sigma_{\ast}\sigma_{\bullet}} \nonumber \\
&& \times \frac{L_{\rm tot}}{L_{\rm tot}-10^{x}}\exp{[-(x-\log \bar{L_{\ast}})^{2}/2\sigma_{\ast}^2]}\nonumber \\
&& \times \exp{[-[\log (L_{\rm tot}-10^{x})-\bar{L_{\bullet}}]^{2}/2\sigma_{\bullet}^2]}
\end{eqnarray}
where $\bar{L_{\ast}} = \bar{L_{\ast}}(z\vert M_{\rm vir},z_{\rm vir})$ and $\bar{L_{\bullet}} = \bar{L_{\bullet}}(z\vert M_{\rm vir},z_{\rm vir})$ are the redshift-dependent mean starburst and AGN luminosities at given halo mass, $M_{\rm vir}$, and virialization redshift, $z_{\rm vir}$, respectively, and $\sigma_{\ast}$ and $\sigma_{\bullet}$ are the corresponding dispersions. $dN_{ST}/dt$ is the dark matter halo mass function \citep{ST99}.

The conditional luminosity function of the starburst component, $\Phi_{\ast}(L_{\ast}\vert L_{\rm tot},z)$, i.e. the number density of sources at redshift $z$ with starburst luminosity $L_{\ast}$ given that the total luminosity is $L_{\rm tot}$, writes
\begin{eqnarray}\label{eq:prob4}
&& \Phi_{\ast}(L_{\ast}\vert L_{\rm tot},z)d\log L_{\ast} = \int_{M{\rm min}}^{M_{\rm max}} dM_{\rm vir} \nonumber \\
&& \times \int_{z}^{z_{\rm max}} dz_{\rm vir} \bigg\vert \frac{dt_{\rm vir}}{dz_{\rm vir}} \bigg\vert \frac{dN_{ST}}{dt_{\rm vir}} (M_{\rm vir},z_{\rm vir}) \frac{d\log L_{\ast}}{2\pi\sigma_{\ast}\sigma_{\bullet}} \nonumber \\
&& \times \frac{L_{\rm tot}}{L_{\rm tot}-L_{\ast}}\exp{[-(\log L_{\ast}-\log \bar{L_{\ast}})^{2}/2\sigma_{\ast}^2]}\nonumber \\
&& \times \exp{[-[\log (L_{\rm tot}-L_{\ast})-\log \bar{L_{\bullet}}]^{2}/2\sigma_{\bullet}^2]},
\end{eqnarray}
while the conditional luminosity function of the AGN component writes
\begin{eqnarray}\label{eq:prob6}
&& \Phi_{\bullet}(L_{\bullet}\vert L_{\rm tot},z)d\log L_{\bullet} = \int_{M{\rm min}}^{M_{\rm max}} dM_{\rm vir} \nonumber \\
&& \times \int_{z}^{z_{\rm max}} dz_{\rm vir} \bigg\vert \frac{dt_{\rm vir}}{dz_{\rm vir}} \bigg\vert \frac{dN_{ST}}{dt_{\rm vir}} (M_{\rm vir},z_{\rm vir}) \frac{d\log L_{\bullet}}{2\pi\sigma_{\ast}\sigma_{\bullet}} \nonumber \\
&& \times \frac{L_{\rm tot}}{L_{\rm tot}-L_{\bullet}}\exp{[-[\log (L_{\rm tot}-L_{\bullet})-\log \bar{L_{\ast}}]^{2}/2\sigma_{\ast}^2]}\nonumber \\
&& \times \exp{[-[\log (L_{\bullet}-\log \bar{L_{\bullet}}]^{2}/2\sigma_{\bullet}^2]}.
\end{eqnarray}
The probability that an object at redshift $z$ has a starburst luminosity $L_{\ast}$ or an AGN luminosity $L_{\bullet}$ given the total luminosity of $L_{\rm tot}=L_{\ast}+L_{\bullet}$ is then:
\begin{equation}\label{eq:prob1}
P_{\ast}(L_{\ast}\vert L_{\rm tot},z)d\log L_{\ast} \equiv \frac{\Phi_{\ast}(L_{\ast}\vert L_{\rm tot},z)d\log L_{\ast}}{\Phi(L_{\rm tot},z)},
 \end{equation}
or
\begin{equation}\label{eq:prob5}
P_{\bullet}(L_{\bullet}\vert L_{\rm tot},z)d\log L_{\bullet} \equiv \frac{\Phi_{\bullet}(L_{\bullet}\vert L_{\rm tot},z)d\log L_{\bullet}}{\Phi(L_{\rm tot},z)}.
\end{equation}
The two probabilities have unit integral over the luminosity range 0--$L_{\rm tot}$ and are linked by the condition that, at fixed $L_{\rm tot}$, $P_{\ast}(L_{\ast}\vert L_{\rm tot}) =-P_{\bullet}(L_{\bullet}\vert L_{\rm tot})\,(d\log L_{\bullet}/d\log L_{\ast})$ with $d\log L_{\bullet}/d\log L_{\ast}= d\log (L_{\rm tot}-L_{\ast})/d\log L_{\ast}=- L_{\ast}/(L_{\rm tot}-L_{\ast})$. Then
\begin{equation}\label{eq:prob7}
P_{\bullet}(L_{\bullet}\vert L_{\rm tot}) = \frac{L_{\bullet}}{L_{\rm tot}-L_{\bullet}}P_{\ast}(L_{\rm tot}-L_{\bullet}\vert L_{\rm tot})
\end{equation}
or
\begin{equation}\label{eq:prob8}
P_{\ast}(L_{\ast}\vert L_{\rm tot}) = \frac{L_{\ast}}{L_{\rm tot}-L_{\ast}}P_{\bullet}(L_{\rm tot}-L_{\ast}\vert L_{\rm tot}).
\end{equation}
Each proto-spheroid of luminosity $L_{\rm tot}$ was assigned bolometric luminosities of the starburst and of the AGN components drawn at random from the above probability distributions, with the parameter values given by \citet{Cai13}. Note that for the starburst component the bolometric luminosity is very close to the IR (8--$1000\,\mu$m) luminosity, but this is not the case for AGNs, especially for type-1's. For example, adopting the AGN SEDs used by \citet{Cai13} we get $L_{\rm IR}/L_{\rm bol} = 0.189$ and 0.776 for type-1's, type-2's, respectively. This should be taken into account when comparing our results with those by \citet{Spin12} which rely on correlations between line and IR luminosity rather than between line and bolometric luminosity, as done in this paper.

%luminosity of the starburst component is very close to \textbf{its} bolometric luminosity. In the case of AGNs, however, the IR to bolometric luminosity ratio varies substantially with type. Adopting the AGN SEDs used by \citet{Cai13} we get $L_{\rm IR}/L_{\rm bol} = 0.189$ and 0.776 for type-1's, type-2's, respectively. %For AGNs associated to proto-spheroids, named type-3 by \citet{Cai13}, for which a strong contribution to the obscuration of the nucleus comes from the interstellar medium of the host galaxy, we get $L_{\rm IR}/L_{\rm bol} = 0.92$. In the following we will investigate the correlations between line and AGN bolometric luminosity.

The model makes detailed predictions also for strongly gravitationally lensed galaxies. These objects are of special interest because strong lensing allows us to measure their total mass distribution up to very large distances and to gain information on sources too faint to be detected with current instrument sensitivities, thus testing models for galaxy formation and dark matter. To deal with them the probability distributions have been applied to the de-lensed luminosities. The de-lensing was made attributing to each lensed source an amplification factor randomly extracted from the amplification probability distributions by \citet{Neg07} and \citet{Lapi12}. After having done that, the luminosities of both component were re-amplified by the same factor.

As mentioned above, the evolution of late-type galaxies, the dominant star forming population at $z\lsim 1.5$, is described by a phenomenological, parametric model, distinguishing two sub-populations, ``cold'' (normal) and ``warm'' (starburst) galaxies. The AGNs associated to these populations are evolved independently, again with a phenomenological recipe. To combine galaxies and AGNs into sources including both components we have exploited the correlation between SFR and average black hole accretion rate derived by \citet{Chen13} on the basis of a sample of 1767 far-IR selected galaxies in the redshift range $0.25 < z < 0.8$. Using their eq.~(5), their bolometric correction for the AGN X-ray emission $L_{\rm bol}/L_{\rm X}=22.4$, and their relationship between the IR luminosity of the starburst, $L_{\rm IR}$, and the SFR we got $\langle L_{\rm bol}/L_{\rm IR}\rangle=0.054(L_{\rm IR}/10^{12}\,L_\odot)^{0.05}$.

It is also necessary to properly take into account the large (see the discussion in Sect.~\ref{sect:intro}) dispersion around the mean relation. Such dispersion is difficult to evaluate because of the large fraction of galaxies in the \citet{Chen13} sample for which only upper limits to the nuclear emission are available. We have estimated it by trial and error, looking for a  value that yields redshift-dependent AGN luminosity functions consistent with the observational estimates by \citet{Delvecchio14}.  In practice, we have carried out simulations associating to galaxies with a given $L_{\rm IR}$ an AGN with bolometric luminosity drawn at random from a Gaussian distribution with mean given by the average $L_{\rm bol}/L_{\rm IR}$ relation and several values of the dispersion. The contribution of optically selected AGNs (see below) was then added to the derived luminosity functions. We get consistency with the data (see a couple of examples in Fig.\,\ref{fig:LF_bol}) with a dispersion of 0.69 dex, in agreement with the distribution of data points in Figs~3 and 4 of \citet{Chen13}. The fact that the model appears to over-estimate the low-luminosity tail of the AGN bolometric luminosity function is not worrisome for the present purposes since these sources are anyway too weak to be detectable by SPICA/SAFARI. It should also be noticed that the observational errors are purely statistical, i.e. do not include uncertainties on the decomposition of the observed spectral energy distribution to disentangle the AGN contribution from that of the host galaxy, on the bolometric correction, on the correction for incompleteness. These systematic errors increase with decreasing AGN luminosity. The evolution of sources as a whole is then driven by that of the SFR, as modeled by \citet{Cai13}.

Not all AGNs obey the accretion rate -- SFR correlation. In particular, hosts of bright optically selected QSOs, with high accretion rates, are found to have SFRs ranging from very low to moderate, particularly at relatively low redshifts  \citep{Schweitzer06,Netzer2007}. These objects were taken into account adopting the best fit evolutionary model by \citet{Croom2009} up to $z=2$ since at higher redshifts the abundance of AGNs associated to late-type galaxies is negligible compared to that of AGNs associated to proto-spheroids whose evolution up to the optically bright phase is described by the  \citet{Cai13} physical model.

The consistency between the model and the observed bolometric luminosity functions, illustrated by Fig.\,\ref{fig:LF_bol} for $z\le 1.5$ and demostrated by \citet{Cai13} at higher $z$, is crucial for the reliability of the derived line luminosity functions because they are calculated coupling the bolometric luminosity functions with the relations between line and bolometric luminosities (see Sect.~\ref{sect:line_vs_IR}). 

In summary, the different IR galaxy populations considered in our paper are:
\begin{itemize}
\item proto-spheroidal galaxies (unlensed and lensed), the progenitors of early-type galaxies and the dominant star forming population at $z\gsim 1.5$, whose evolution is described by the physical model by \citet{Cai13};
\item AGNs associated to proto-spheroids whose co-evolution with the host galaxies is self-consistently built in the \citet{Cai13} physical model;
\item late-type galaxies, the dominant star forming population at $z\lsim 1.5$, divided into two sub-populations (spiral and starburst galaxies), whose evolution is described by the \citet{Cai13} phenomenological model;
\item type-1 and type-2 AGNs associated to late-type galaxies, whose evolution is linked to that of host galaxies via the \citet{Chen13} accretion rate -- SFR correlation;
\item optical AGNs, whose evolution is described by the \citet{Croom2009} best-fit model up to $z=2$ and included in the \citet{Cai13} physical model at higher $z$.
\end{itemize}
While referring to the \citet{Cai13} paper for full details, we summarize here, for the sake of clarity, a few points. Proto-spheroidal galaxies and massive bulges of disk galaxies are assumed to \textit{virialize} at $z\ge 1.5$ but their evolution is followed down to $z=0$. Their lifetimes in the star formation phase increase with decreasing halo mass, consistent with the observed ``downsizing'', from $\simeq 0.7\,$Gyr for the most massive ones to a few Gyr for the least massive. Hence they are contributing to the IR luminosity functions also at $z<1.5$, but their contribution decreases with decreasing $z$. At $z< 1.5$ late-type galaxies take over them as the dominant contributors. The phenomenological model describing their evolution includes a smooth decline of their space density above $z\simeq 1.5$ but their contribution to the global IR luminosity function is still significant up to $z\simeq 2$. 

The evolution of supermassive black holes associated to proto-spheroidal galaxies is followed by the physical model \textit{only} during their active star formation phase when black holes acquire most of their mass. In other words the present implementation of the physical model is unable to follow the later AGN evolution when spheroids are in essentially passive evolution and nuclei can be reactivated by, e.g., interactions, mergers, or dynamical instabilities, especially if they acquire a disk, i.e. become the bulges of later-type galaxies. These later phases are dealt with by the phenomenological model, as described above. 

%\begin{figure*}
%\includegraphics[trim=0.7cm 3.7cm 1.1cm 4.0cm,clip=true,width=\textwidth]{LF_IR_hybrid_sb_sp_effect.eps} %\vspace{-0.5cm}
%\caption{At three different redshifts, IR (8-1000\,$\mu$m) luminosity function for the late-type population of galaxies, constructed from the \citet{Cai13} model, showing the contribution of the AGN emission computed using the \citet{Chen13} relation (as explained in the text).} FIGURA ELIMINATA PERCHE' PUO' ESSERE "PERICOLOSA": NON E' GARANTITO CHE LE FUNZIONI DI LUMINOSITA' DEGLI AGN BASATE SULLA NOSTRA RICETTA SIANO CONSISTENTI CON I DATI. AD OGNI MODO PROBABILMENTE NON POSSIAMO FARE MEGLIO DI COSI'.
%\label{fig:LF_IR_hybrid_sb_sp_effect}
%\end{figure*}

\begin{figure*}
\includegraphics[trim=0.6cm 2.4cm 1.2cm 3.0cm,clip=true,width=\textwidth]{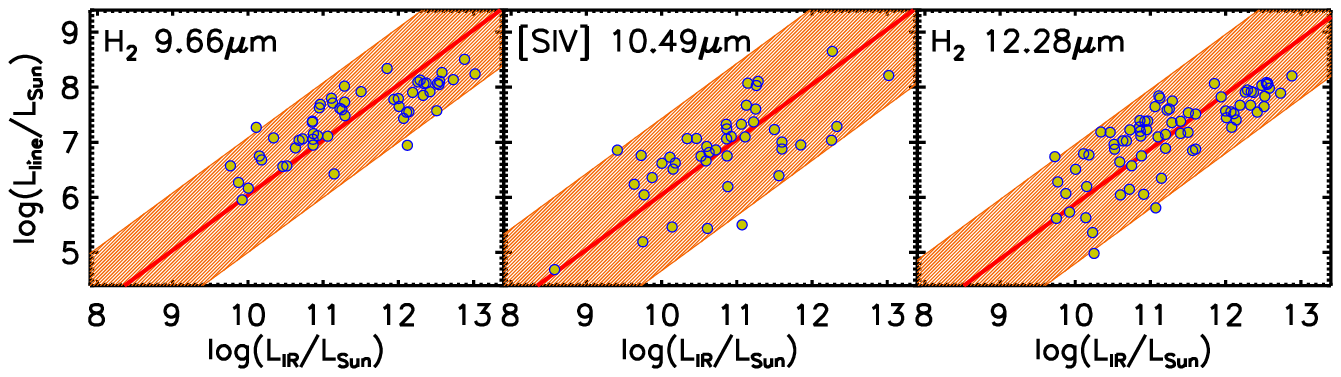} %\vspace{-0.5cm}
\caption{Line luminosity as a function of the IR luminosity of the starburst component for the ${\rm (H_{2})}9.66\mu$m, ${\rm [SIV]}10.49\mu$m and ${\rm (H_{2})}12.28\mu$m lines. Luminosities are in solar units. Data from \citet{Roussel06,Farrah07,Bern09,Hao09,O'Dowd09,O'Dowd11,Pereira-Santaella10,Cormier12}, excluding all objects for which there is evidence of a substantial AGN contribution. The orange bands show the $2\,\sigma$ spread around the mean linear relations (red lines) in the $\log$--$\log$ plane. The mean ratios $\langle \log(L_{\ell}/L_{\rm IR})\rangle$ and the dispersions around them are given in Table\,\ref{tab:sb_c_values}.}
 \label{fig:sb_line_vs_IR}
\end{figure*}

%%%%%%%
% TABLE %
%%%%%%%%

%%%%%%%%
\begin{table}%[h]
\centering
\footnotesize
\begin{tabular}{lc}
\hline
\hline
\rule[-3mm]{0mm}{6mm}
Spectral line & $\displaystyle\big\langle\log\big(\displaystyle{L_{\ell}\over L_{\rm IR}}\big)\big\rangle\ (\sigma)$ \\
\hline
${\rm PAH 6.2}\mu$m & -2.20\ (0.36)  \\
${\rm PAH 7.7}\mu$m & -1.64\ (0.36)  \\
${\rm PAH 8.6}\mu$m & -2.16\ (0.36)  \\
${\rm H_{2}}9.66\mu$m & -3.96\ (0.52)  \\
${\rm [SIV]}10.49\mu$m & -3.95\ (0.69)  \\
${\rm H_{2}}12.28\mu$m & -4.12\ (0.54)  \\
${\rm [NeII]}12.81\mu$m & -3.11\ (0.45)$^{1}$  \\
${\rm [NeIII]}15.55\mu$m & -3.69\ (0.47)$^{1}$  \\
${\rm H_{2}}17.03\mu$m & -4.04\ (0.46)$^{1}$ \\
${\rm [SIII]}18.71\mu$m & -3.49\ (0.48)$^{1}$  \\
${\rm [SIII]}33.48\mu$m & -3.05\ (0.31)$^{1}$  \\
\hline
\multicolumn{2}{l}{\scriptsize{$^{1}$Taken from \citet{Bonato2014}}}\\
\hline
\hline
\end{tabular}
\caption{Mean values of the log of line to IR (8-1000\,$\mu$m) continuum luminosities for star forming galaxies, $\langle\log({L_{\ell}/L_{\rm IR}})\rangle$, and associated dispersions $\sigma$. For the H$_{2}\, 17.03\,\mu$m line, the tabulated values, taken from \citet{Bonato2014}, have been computed excluding local ULIRGs, for which the luminosity in this line was found to be uncorrelated with $L_{\rm IR}$. The mean value of $\log(L_{\ell})$ in this line for local ULIRGs was found to be $\log(L_{\ell}/L_\odot)=8.07$ with a dispersion of 0.34.}
\label{tab:sb_c_values}
\end{table}

\section{Line versus IR luminosity}\label{sect:line_vs_IR}
%%%%%%%%%%%%%%%%%%%%%%%%%%%%%%%%%%%%%%%%%%%%%%%%%%%%%%%%%%%%%%%%%%%%%%%%%%%%%

IR lines excited by star formation activity have been dealt with by \citet{Bonato2014}, with an analysis of the relations between line and starburst IR luminosities that considered differences among source populations and that was supported by extensive simulations taking into account dust obscuration. We complement their study considering the brightest lines that are excited by, or also by, AGNs. These include 3 typical AGN lines ([NeV]\-14.32\,$\mu$m, [NeV]\-24.31\,$\mu$m and [OIV]\-25.89\,$\mu$m) and 8 lines that can also be produced in star formation regions (the 5 HII region lines: [SIV]\-10.49\,$\mu$m, [NeII]\-12.81\,$\mu$m, [NeIII]\-15.55\,$\mu$m, [SIII]\-18.71\,$\mu$m and [SIII]\-33.48\,$\mu$m; three molecular hydrogen lines, H$_{2}$\,\-9.66\,$\mu$m, H$_{2}$\,\-12.28\,$\mu$m and H$_{2}$\,\-17.03\,$\mu$m). The other IR lines considered by \citet{Bonato2014} have negligible AGN contributions so that we adopt directly the luminosity functions derived in that paper. In addition we present results for the $6.2$, $7.7$ and $8.6\mu$m  Polycyclic Aromatic Hydrocarbons (PAH) lines, overlooked by \citet{Bonato2014} but of considerable interest for SPICA/SAFARI surveys. Of course, to detect the broad PAH lines the SAFARI ($R=2000$) spectral resolution needs to be degraded without loss of signal, using appropriate algorithms. All the lines have been taken into account to compute the number of sources detected in more than one line.

As in \citet{Bonato2014}, the line luminosities are derived from the continuum ones exploiting the relations between line and IR (or bolometric in the AGN case) luminosities, taking into account the dispersions around the mean values. For 5 of the lines listed above, namely [NeII]\-12.81\,$\mu$m, [NeIII]\-15.55\,$\mu$m, H$_{2}$\,\-17.03\,$\mu$m, [SIII]\-18.71\,$\mu$m and [SIII]\-33.48\,$\mu$m, these relationships were already derived, for the starburst component only, by  \citet{Bonato2014}. Again for the starburst component we have collected from the literature data on 3 additional lines (H$_{2}$\,\-9.66\,$\mu$m, [SIV]\-10.49\,$\mu$m and H$_{2}$\,\-12.28\,$\mu$m), excluding objects for which there is evidence of a substantial AGN contribution. These data are shown in Fig.~\ref{fig:sb_line_vs_IR} together with the best fit linear relations. We have neglected the starburst contribution to the 3 typical AGN lines.

The mean ratios $\langle \log(L_{\ell}/L_{\rm IR})\rangle$ for star forming galaxies and the dispersions around them are given in Table\,\ref{tab:sb_c_values}. The values for the $6.2$, $7.7$ and $8.6\mu$m PAH lines (also given in Table\,\ref{tab:sb_c_values}) were inferred from that for the PAH\,$11.25\,\mu$m line, given by \citet{Bonato2014}, using the average ratios $L_{PAH_{6.2}}/L_{PAH_{11.25}}=1.22$, $L_{PAH_{7.7}}/L_{PAH_{11.25}}=4.5$ and $L_{PAH_{8.6}}/L_{PAH_{11.25}}=1.36$ given by \citet{Fiolet10}.

The relationships between line and bolometric luminosities of AGNs were determined using the data on the QSO samples by \citet{Veilleux09}, on the local Seyfert galaxies whose mid-IR emission was found to be 100\% AGN dominated by \citet{Sturm02} and, only for the three pure AGN lines ([NeV]\-14.32\,$\mu$m, [NeV]\-24.31\,$\mu$m and [OIV]\-25.89\,$\mu$m), on the local Seyfert galaxies by \citet{Tomm08,Tomm10} coupled with the continuum nuclear emission measurements by \citet{Gorjian04}, \citet{Gandhi09} and \citet{Honig10}.

The IR luminosities given in \citet{Veilleux09} and \citet{Sturm02} are defined as in \citet{SandersMirabel1996}, but this definition applies to starburst not to AGN SEDs. Therefore to convert them to AGN bolometric luminosities we cannot use the correction factors given in Sect.~\ref{sect:evol}. The AGN bolometric luminosities were then estimated using the appropriate \citet{Cai13} SEDs (i.e. for QSOs and Seyfert 1 galaxies the mean QSO SED of \citealt{Richards06} extended to millimeter wavelengths as described in \citealt{Cai13}; for Seyfert 2 galaxies the SED of the local AGN-dominated ULIRG Mrk 231 taken from the SWIRE library) normalized to the observed $12\,\mu$m luminosities which are expected to be modestly affected by contamination from the host galaxy. Normalizing instead to the AGN $6\,\mu$m fluxes of \citet{Sani10}, we obtain, for a sub-sample containing 65\% of our objects, an average decrease of the estimated AGN bolometric luminosity by about 17\%. On the other hand we caution  that the Mrk 231 SED might have a non-negligible host galaxy contamination. In fact its far-IR emission peaks at $\sim 100 \,\mu$m. At such relatively long wavelength the contribution from dust heated by young stellar populations can be substantial or even dominant  (\citealt{Mullaney11}). Therefore  our approach likely overestimates the bolometric luminosities of Seyfert 2 galaxies. This however has little impact on our conclusions since these AGNs are minor contributors to line counts at the SPICA/SAFARI detection limits.

\begin{figure*}
\includegraphics[trim=0.6cm 0.7cm 1.4cm 1.1cm,clip=true,width=\textwidth]{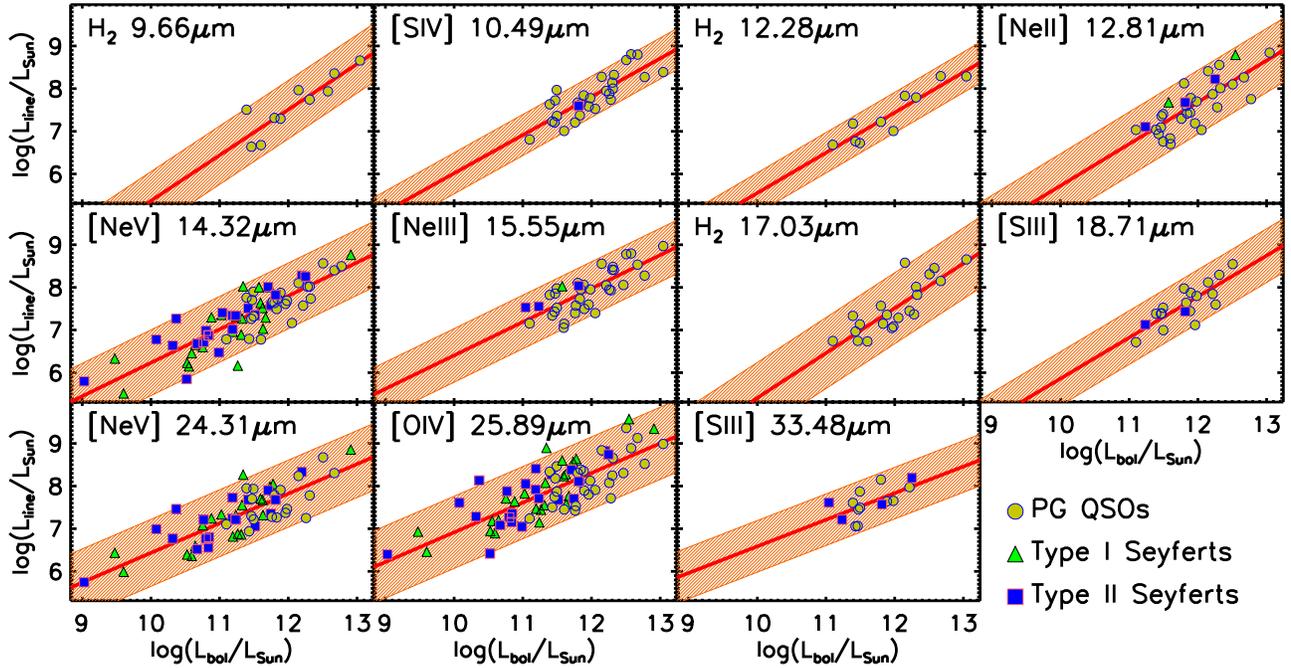} %\vspace{-0.5cm}
\caption{Line versus bolometric luminosity, in solar units, for AGNs. {PG QSO data (yellow circles) are taken from \citet{Veilleux09}, Seyfert 1 (green triangles) and Seyfert 2 (blue squares) data from \citet{Sturm02} and, only for the [NeV]\-14.32\,$\mu$m, [NeV]\-24.31\,$\mu$m and [OIV]\-25.89\,$\mu$m lines, also from \citet{Tomm08,Tomm10}. The orange bands show the $2\,\sigma$ spread around the linear relations $\log(L_{\ell})=a\cdot\log(L_{\rm bol}) + b$ (red line).}}
 \label{fig:line_vs_IR}
\end{figure*}

At variance with star forming galaxies for which data and extensive simulations are consistent with a constant line-to-continuum IR luminosity ratio \citep{Bonato2014}, for AGNs observational data indicate luminosity dependent ratios \citep{Hill13}. We adopt a relation of the form $\log({L_{\ell}})=a\cdot\log({L_{\rm bol}})+b$. The best fit coefficients for the 11 lines of our sample and the dispersions, $\sigma$, are listed in Table~\ref{tab:agn_a_b_values}. There are no significant differences among the line--bolometric luminosity correlations for the different AGN types (see Fig.\,\ref{fig:line_vs_IR}).

\begin{table}%[h]
\centering
\footnotesize
\begin{tabular}{lccc}
\hline
\hline
\rule[-3mm]{0mm}{6mm}
Spectral line & $a$ & $b$ & disp ($1\sigma$)\\
\hline
${\rm H_{2}}9.66\mu$m       & 1.07 & -5.32 & 0.34 \\
${\rm [SIV]}10.49\mu$m      & 0.90 & -2.96 & 0.24 \\
${\rm H_{2}}12.28\mu$m      & 0.94 & -3.88 & 0.24 \\
${\rm [NeII]}12.81\mu$m     & 0.98 & -4.06 & 0.37 \\
${\rm [NeV]}14.32\mu$m      & 0.78 & -1.61 & 0.39 \\
${\rm [NeIII]}15.55\mu$m    & 0.78 & -1.44 & 0.31 \\
${\rm H_{2}}17.03\mu$m      & 1.05 & -5.10 & 0.42 \\
${\rm [SIII]}18.71\mu$m     & 0.96 & -3.75 & 0.31 \\
${\rm [NeV]}24.31\mu$m      & 0.69 & -0.50 & 0.39 \\
${\rm [OIV]}25.89\mu$m      & 0.70 & -0.04 & 0.42 \\
${\rm [SIII]}33.48\mu$m     & 0.62 &  0.35 & 0.30 \\
\hline
\hline
\end{tabular}
\caption{Coefficients of the best-fit linear relations between line and AGN bolometric luminosities, $\log({L_{\ell}})=a\cdot\log({L_{\rm bol}})+b$, and $1\sigma$ dispersions associated to the relations.}
\label{tab:agn_a_b_values}
\end{table}

%%%%%%%%%%%%%%%%%%%%%%%%%%%%%%%%%%%
\section{Line luminosity functions and number counts}\label{sect:LF}
%%%%%%%%%%%%%%%%%%%%%%%%%%%%%%%%%%%

\begin{figure*}
\makebox[\textwidth][c]{
\includegraphics[trim=0.0cm 2.5cm 0.6cm 2.6cm,clip=true,width=\textwidth, angle=0]{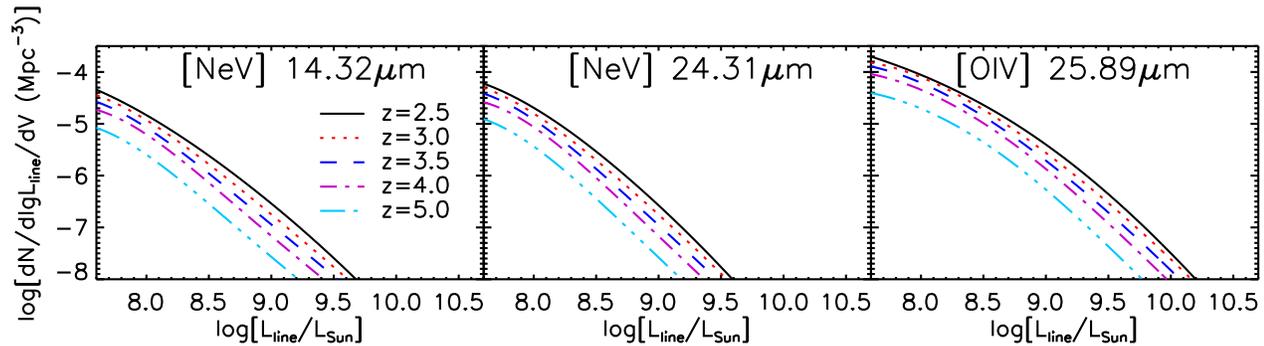}
}
\caption{Predicted luminosity functions of the three typical AGN lines of our sample at different redshifts at  which these lines can be detected by SPICA/SAFARI.}
 \label{fig:LF_agn}
\end{figure*}

\begin{figure*}
\makebox[\textwidth][c]{
\includegraphics[trim=0.0cm 1.2cm 0.6cm 1.6cm,clip=true,width=\textwidth, angle=0]{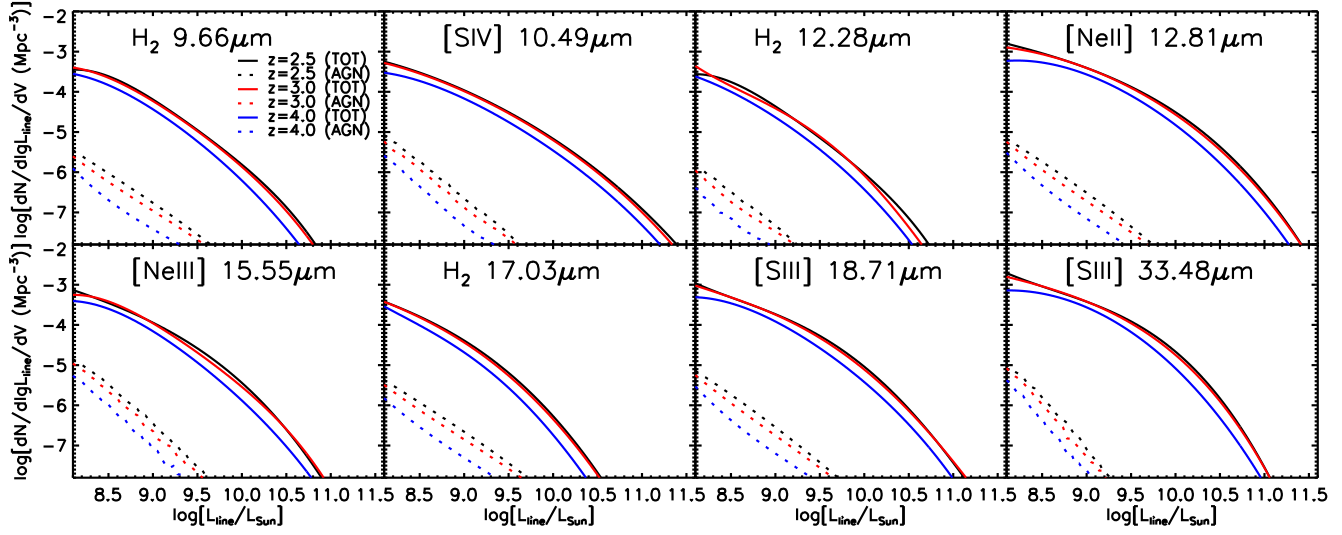}
}
\caption{Predicted luminosity functions for the total (starburst plus AGN) emission (solid lines) and for the AGN component only (dashed lines) at different redshifts at which these lines can be detected by SPICA/SAFARI.}
 \label{fig:LF_tot}
\end{figure*}

To compute the line luminosity functions and the number counts we adopted a Monte Carlo approach, starting from the redshift-dependent luminosity functions of the source populations described in Sect.~\ref{sect:evol}, including both the starburst and the AGN component. Each source was assigned luminosities of each component drawn at random from the appropriate probability  distributions (see Sect.~\ref{sect:evol}).
%
%The simulation is performed as follows. For a given galaxy population and a given redshift bin, $\Delta z$, we draw from the corresponding IR luminosity function a number of IR luminosities equal to that of the objects expected over a (reference) area $\Omega=0.5$\,deg$^{2}$, and with IR luminosity greater than a minimum value that we set\footnote{This value is chosen to ensure that all the galaxies detectable by SAFARI are represented in the simulation while keeping the computing time, that depends on the total number of simulated sources, to an acceptable level.} to $\log(L_{\rm IR, min}/L_{\odot})=8.0$, i.e.
%\begin{equation}
%\label{eq:flux_function}
%\mathcal{N}(z,\Omega) = \int_{\log L_{\rm IR, min}}^{+\infty} \!\!\!\!\!\!\Phi(L_{\ell},z)\,{\rm d}\log L_{\rm IR}\, \frac{{\rm d}^2V_{c}(z)}{{\rm d}z\,{\rm d}\Omega}\, \Delta z\, \Omega,
%\end{equation}
%
%where ${\rm d}V_{c}(z)$ is the comoving volume element.
%
Then we associated to each component line luminosities drawn at random from Gaussian distributions with the mean values and the dispersions given in Tables~\ref{tab:sb_c_values} and \ref{tab:agn_a_b_values}. This procedure gives the line luminosity functions of the objects as a whole, as well as of their starburst and AGN components. Examples are shown in Figs \ref{fig:LF_agn} and \ref{fig:LF_tot}.

\begin{figure*}
\hspace{+0.0cm}
\makebox[\textwidth][c]{
\includegraphics[trim=0.0cm -0.5cm 0.4cm 0.0cm,clip=true,width=1.1\textwidth, angle=0]{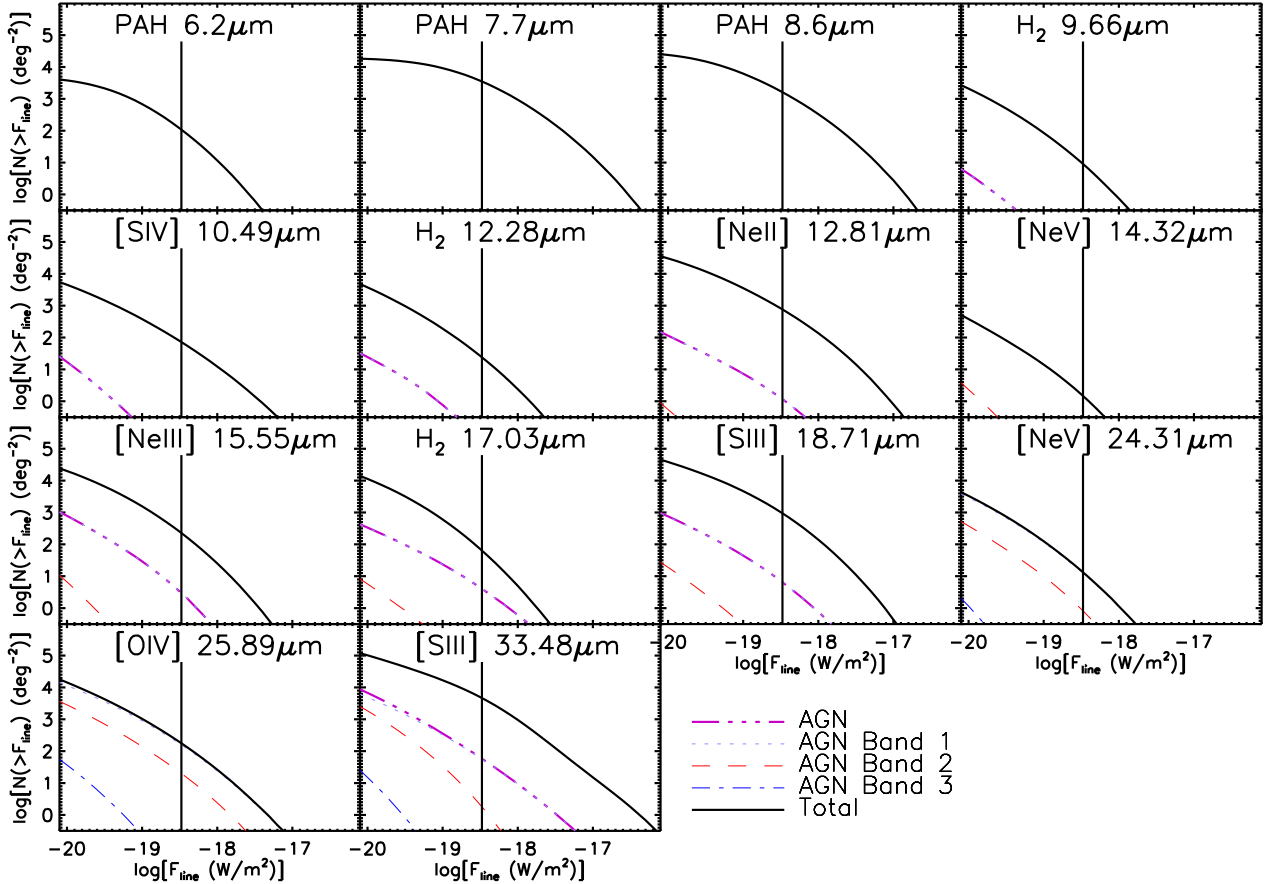}
}
%\vspace{-4.5cm}
\caption{Integral counts in several lines of galaxies as a whole (starburst plus AGN components; solid black lines) and of AGNs only (dash-dotted violet lines) over the full SPICA-SAFARI wavelength range, and, for the AGNs, in each of its three bands. The vertical lines correspond to the average detection limit in the 3 bands for 1\,hr integration/FoV.}
 \label{fig:intcounts_spica_all}
\end{figure*}

%
%We then associate to each simulated IR source a line luminosity by sampling, at random, the distribution of values of $\log(L_{\ell}/L_{\rm IR})$, assumed to be Gaussian with mean values and dispersions given in Table~\ref{tab:c_values}. Similarly we associate a redshift to each object, assuming a uniform probability distribution within $z\pm(\Delta z/2)$. Finally the line flux, $F_{\ell}$, is computed from the simulated redshift, $z_{\rm simul}$, and the simulated line luminosity, $L_{\ell,\rm simul}$, as $F_{\ell,\rm simul}=L_{\ell,\rm simul}/4\pi d^{2}_{\rm L}(z_{\rm simul})$, $d_{L}(z)$ being the luminosity distance.
The counts are then straightforwardly computed. The wavelength range covered by each SPICA/SAFARI band, $[\lambda_{\rm min}, \lambda_{\rm max}]$, and the line rest-frame wavelength, $\lambda_{\ell}$, define the minimum and the maximum redshift within which the line is  detectable in that band, $z_{\ell,\rm min/max}=\lambda_{\rm min/max}/\lambda_{\ell}-1$. The integral counts predicted by our model in the SPICA/SAFARI wavelength range for the 14 IR lines of our sample are shown in Fig.~\ref{fig:intcounts_spica_all}.

%The simulated line luminosities are binned to produce the line luminosity functions at a given redshift. In order to reduce the effect of fluctuations in the brightest luminosity bins, where the statistic is poorest, the whole procedure is repeated 300 times and the simulated luminosity functions are averaged together.

%The predicted number counts and redshift distributions are also derived by averaging over the 300 simulated catalogues.

%%%%%%%%%%%%
% FIGURE %
%%%%%%%%%%%%
%

%%%%%%%%%%%%
% FIGURE %
%%%%%%%%%%%%
%

%%%%%%%%%%%%
% FIGURE %
%%%%%%%%%%%%

%%%%%%%%%%%%
% FIGURE %
%%%%%%%%%%%%
\begin{figure*}
\hspace{+0.0cm}
\makebox[\textwidth][c]{
\includegraphics[trim=0.2cm 2.0cm 1.3cm 2.5cm,clip=true,width=0.9\textwidth, angle=0]{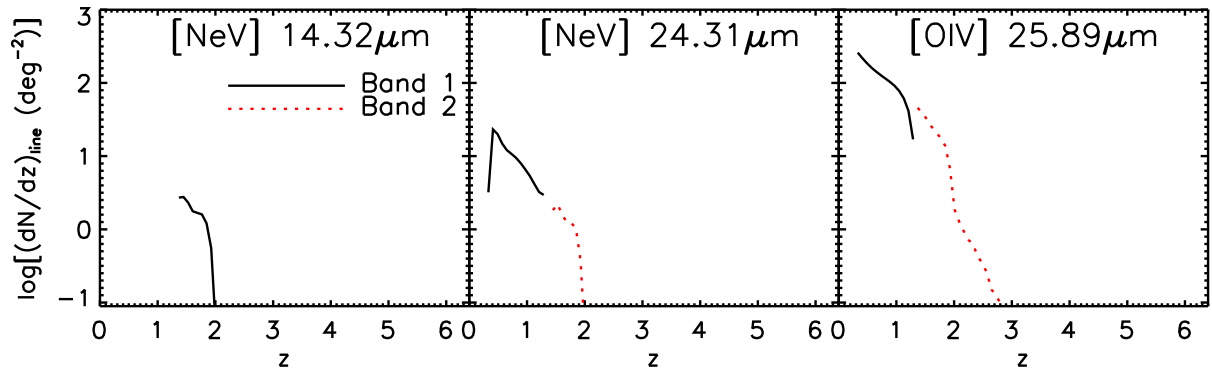}
}
%\vspace{-3.cm}
\caption{Examples of the predicted redshift distributions of AGNs detected by SPICA/SAFARI for a 1\,hr exposure per FoV. The different colours identify the spectral bands.}
 \label{fig:redshiftdistr_spica}
\end{figure*}
%

%%%%%%%%%%%%
% FIGURE %
%%%%%%%%%%%%
\begin{figure*}
\makebox[\textwidth][c]{
\includegraphics[trim=0.4cm 1.7cm 1.0cm 2.5cm,clip=true,width=0.9\textwidth, angle=0]{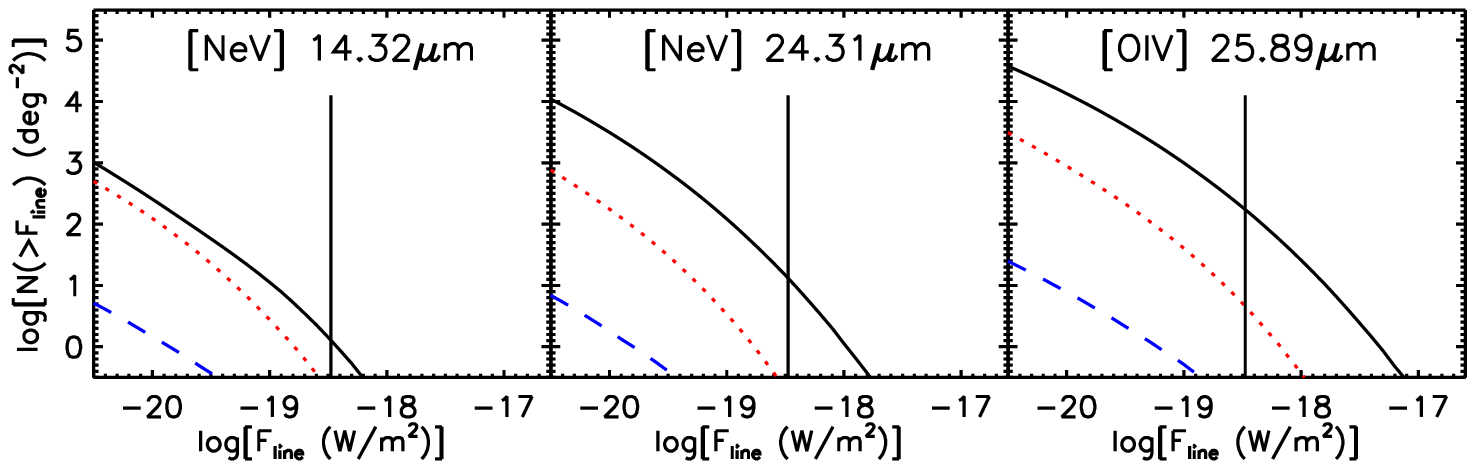}
}
\caption{Contributions of different AGN populations to the integral counts in the [NeV]$14.32\mu$m, [NeV]$24.31\mu$m and [OIV]$25.89\mu$m lines. Solid black line: AGNs associated to late-type galaxies plus optically selected AGNs; dotted red line: AGN associated to unlensed proto-spheroids; dashed blue line: AGNs associated to strongly gravitationally lensed proto-spheroids. The vertical lines correspond to the average value of the detection limits ($3.7\cdot10^{-19}\,\hbox{W}/\hbox{m}^{2}$, $3.4\cdot10^{-19}\,\hbox{W}/\hbox{m}^{2}$, $2.9\cdot10^{-19}\,\hbox{W}/\hbox{m}^{2}$) for  the three bands ($\,34-60\,\mu$m, $\,60-110\,\mu$m, $\,110-210\,\mu$m) for 1-hr exposure per FoV.}
 \label{fig:intcounts_spica}
\end{figure*}

The SPICA/SAFARI reference $5\sigma$ detection limits for an integration of 1\,hr per FoV are (B. Sibthorpe, private communication) $3.7\cdot10^{-19}\,\hbox{W}/\hbox{m}^{2}$ for the first band ($\,34-60\,\mu$m), $3.4\cdot10^{-19}\,\hbox{W}/\hbox{m}^{2}$ for the second band ($\,60-110\,\mu$m) and $2.9\cdot10^{-19}\,\hbox{W}/\hbox{m}^{2}$ for the third band ($\,110-210\,\mu$m).  We present predictions for a survey of 5\,deg$^{2}$ with 1\,hr per FoV. The survey area is 10 times larger compared to the reference survey considered in  \cite{Bonato2014}. In fact, as discussed later on in this section and in Sec.\,6, AGN lines are more difficult to detect than  starburst lines and thus a wider/deeper survey is required to achieve a good statistics.
The predicted numbers of sources detected in each line by a SPICA/SAFARI survey covering $5\,\hbox{deg}^2$ with a 1\,hr integration/FoV are given in the last column of Table~\ref{tab:counts_spica1}. The other columns detail the redshift distributions of detected sources. This SPICA/SAFARI survey will be able to detect many hundreds of AGNs (762, according to our calculations) in the strongest line ([OIV]$25.89\mu$m).

The redshift distributions of sources detected in the three AGN lines [NeV]$14.32\mu$m, [NeV]$24.31\mu$m and [OIV]$25.89\mu$m are displayed in Fig.~\ref{fig:redshiftdistr_spica}, where the redshift ranges covered by the different bands are identified by different colours. Fig.~\ref{fig:intcounts_spica} shows the contributions to the counts of AGNs of the various kinds.

\begin{figure*}
\makebox[\textwidth][c]{
\includegraphics[trim=2.3cm 0.3cm 0.7cm 0.2cm,clip=true,width=0.48\textwidth, angle=0]{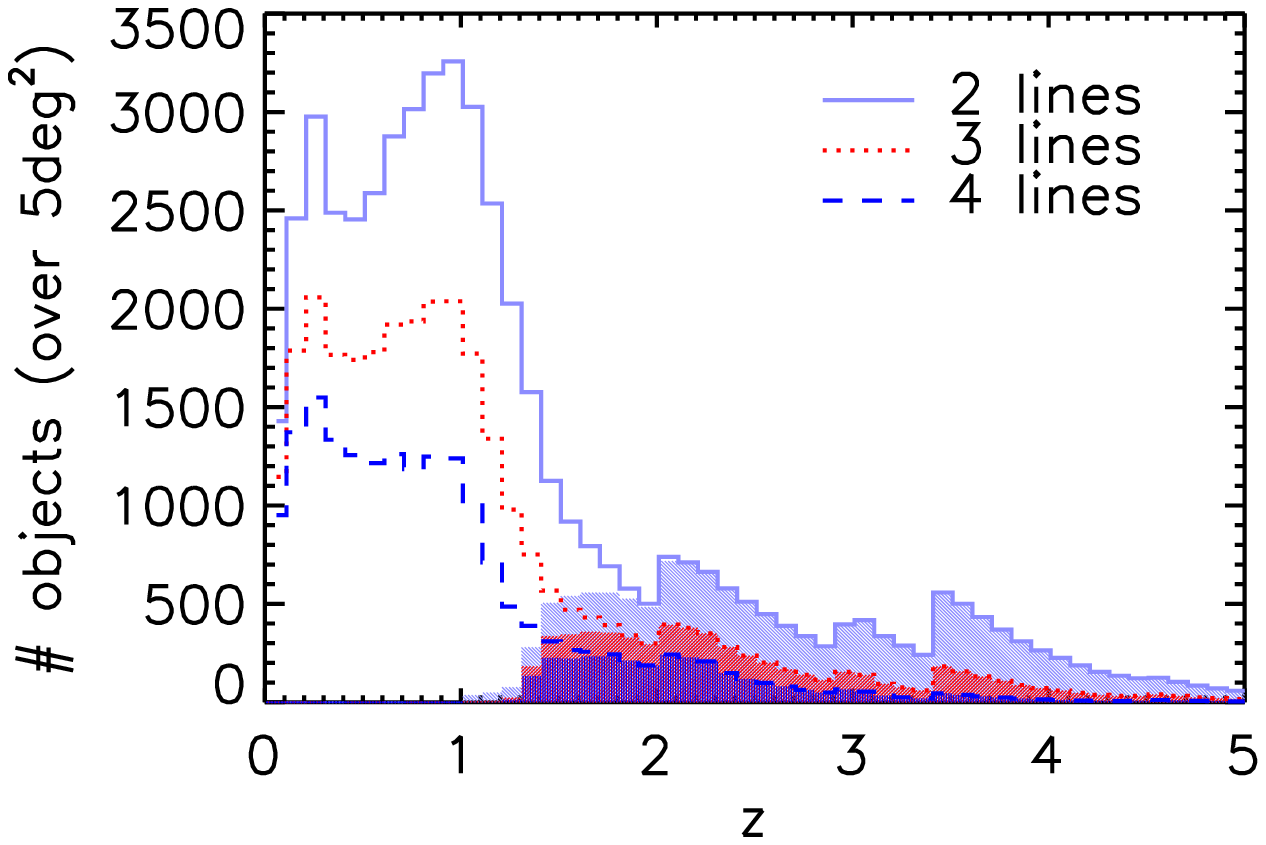}
\includegraphics[trim=2.3cm 0.3cm 0.7cm 0.2cm,clip=true,width=0.48\textwidth, angle=0]{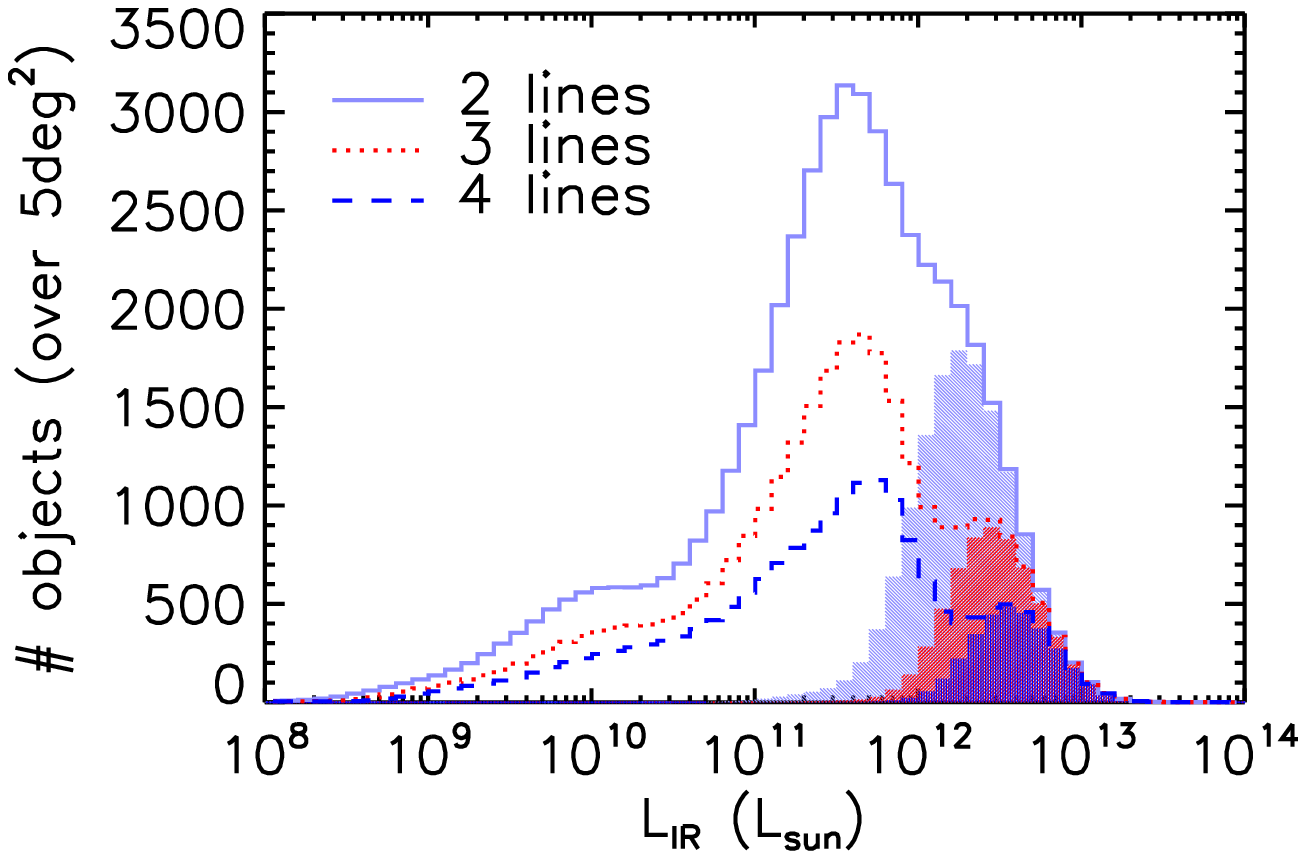}
}
\caption{Predicted redshift (left) and total IR luminosity (right) distributions of galaxies (starburst plus AGN components) detectable in two (cyan histogram), three (red) and four (blue) spectral lines, by a SPICA/SAFARI survey covering $5\,\hbox{deg}^2$ in 1\,hr integration/FoV (taking into account both the spectral lines of this sample and those studied in \citealt{Bonato2014}). The shaded areas represent the contributions of proto-spheroids.}
 \label{fig:z_distr_tot}
\end{figure*}

Fig.~\ref{fig:z_distr_tot} illustrates the luminosity and redshift distributions of sources (starburst plus AGN components) for which the SPICA/SAFARI 5\,deg$^{2}$ survey will detect at least 2, 3 or 4 lines, taking into account both the spectral lines considered in this paper and the pure starburst lines considered in \citet{Bonato2014}. We expect that 114903, 50817, 29492, 18412, 11555 and 6930 sources will be detected in at least 1, 2, 3, 4, 5 and 6 lines, respectively. The numbers of proto-spheroidal galaxies detectable in at least 1, 2, 3, 4, 5 and 6 lines are 44335, 13854, 6149, 3151, 1749 and 907, respectively. Sources detected in at least one line include $\sim$ 724 strongly lensed galaxies at $z>1$; about 300 of them will be detected in at least 2 lines.

\begin{figure*}
\makebox[\textwidth][c]{
\includegraphics[trim=3.2cm 0.7cm 2.2cm 0.2cm,clip=true,width=0.48\textwidth, angle=0]{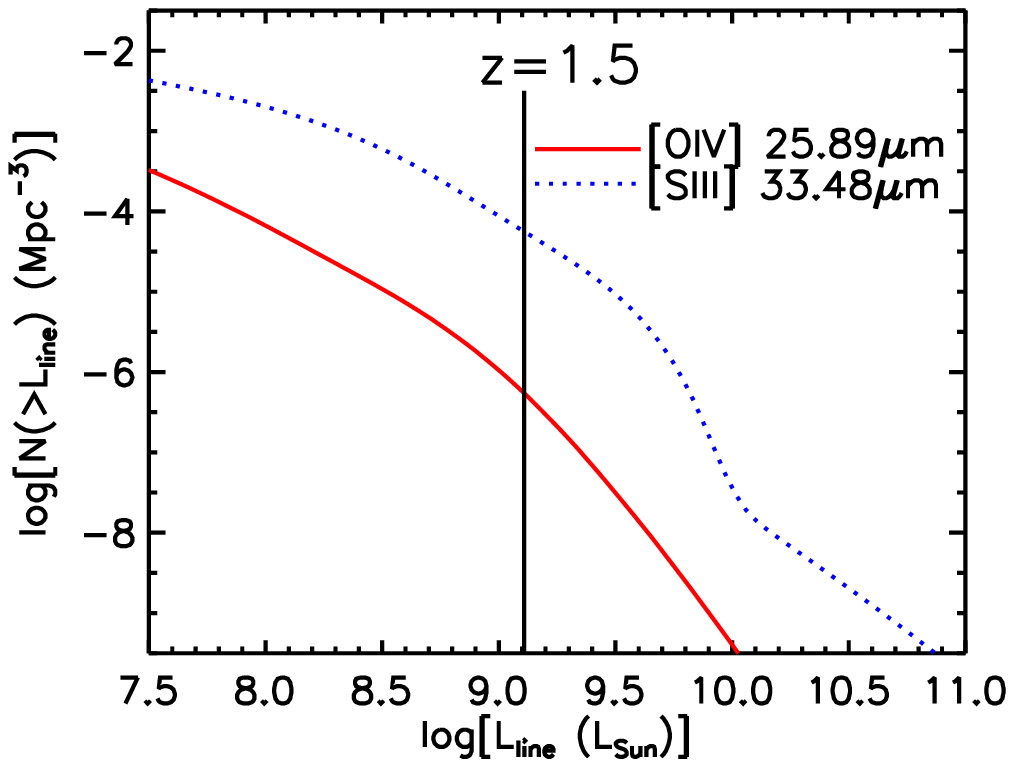}
\includegraphics[trim=3.2cm 0.7cm 2.2cm 0.2cm,clip=true,width=0.48\textwidth, angle=0]{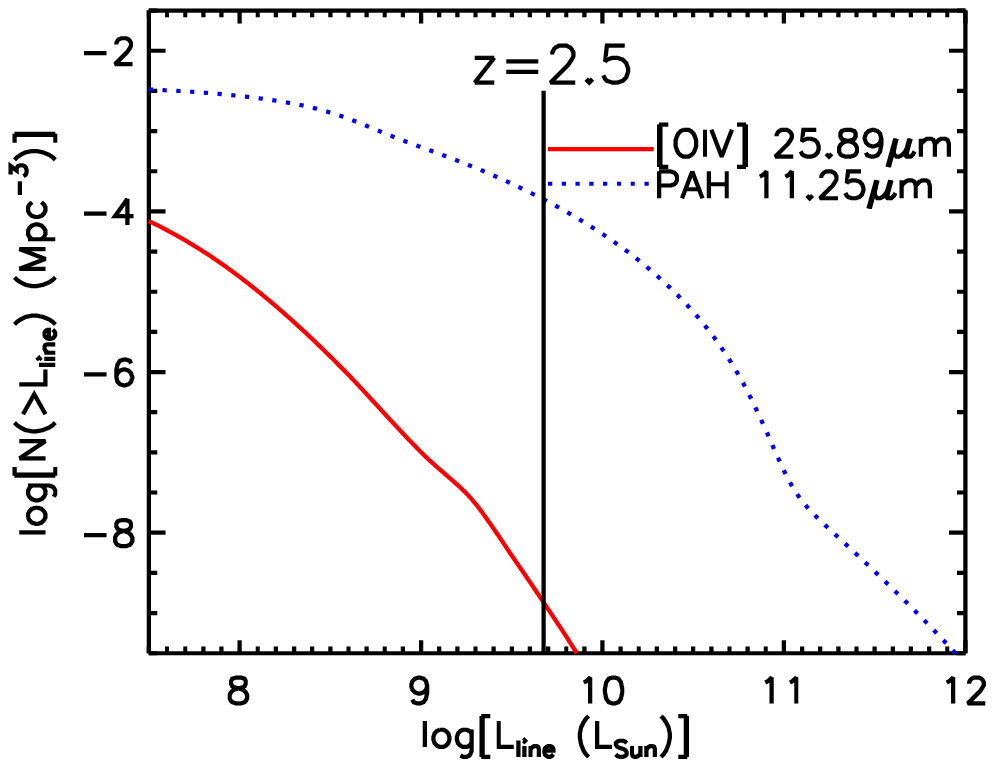}
}
\caption{Comparison of the cumulative luminosity functions at 2 redshifts of the brightest AGN line (${\rm [OIV]}25.89\mu$m) with those of two bright starburst lines. The dotted curves show the contributions of strongly lensed sources  to the total luminosity functions (solid curves). At each redshift the vertical line corresponds to the minimum luminosity detectable by a survey with 1\,hr integration/FoV. }  \label{fig:LFcomp}
\end{figure*}

It is easily seen that these values, once rescaled to the same survey area, a factor of 10 smaller, are similar to those calculated in \citet{Bonato2014} considering only star forming galaxies: the AGN component adds a minor contribution. The reason for that is illustrated by Fig.~\ref{fig:LFcomp} where the luminosity functions of the brightest AGN line (${\rm [OIV]}25.89\mu$m) are compared with those of two bright starburst lines at 2 redshifts. Two factors concur to yield a much higher space density of detectable starbursts compared to  AGNs. First, AGNs are far less numerous than starburst galaxies, due to the much shorter lifetime of their bright phase. Second, many starburst lines have ratios to bolometric luminosity substantially higher than even the brightest AGN line (cf. the luminosity ratios for starbursts in Table~1 of \citealt{Bonato2014} with those for AGNs in Table~\ref{tab:agn_a_b_values} of the present paper).

%%%%%%%%%%%%%%%%%%
\section{Comparison with Previous Estimates}\label{sect:comparison}
%%%%%%%%%%%%%%%%%%

\begin{figure*}
%\hspace{+2.0cm}
%\makebox[\textwidth][c]{
%\includegraphics[width=1.55\textwidth, angle=0]{../LFIR/Grafici_articolo/panel_redshdistr.ps}
\includegraphics[trim=-0.4cm -0.8cm 0.2cm 0.0cm,clip=true,width=\textwidth, angle=0]{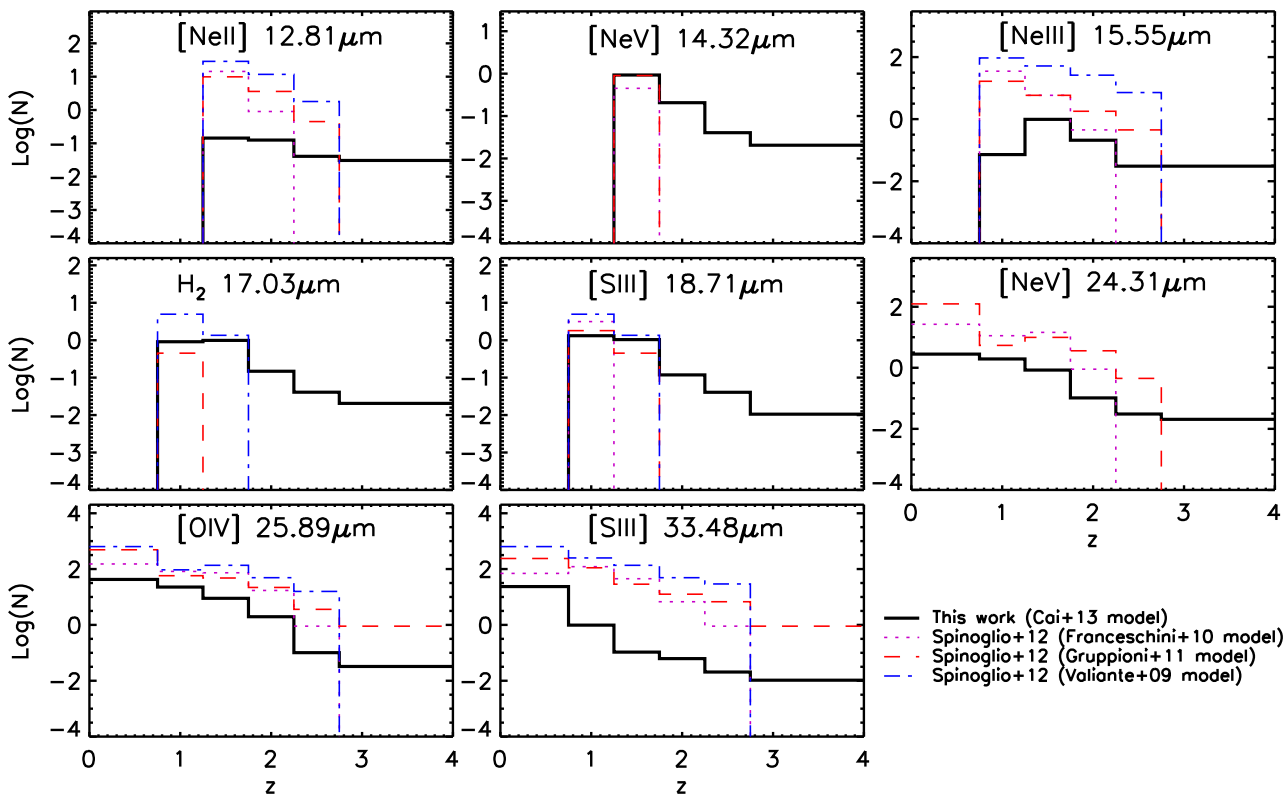}
%}
%\vspace{-2.5cm}
\caption{Redshift distributions of AGNs detectable in each of the 8 lines by a SPICA/SAFARI survey covering $0.5\,\hbox{deg}^2$ with 1\,hr integration/FoV. The predictions of our model (black solid lines) are compared with those of the 3 models used by \citet{Spin12}; see the legend on the bottom right.}
 \label{fig:histo}
\end{figure*}

Fig.~\ref{fig:histo} compares our predictions for the redshift distributions of sources detected by SPICA/SAFARI with those from the 3 models used by \citet{Spin12}. The total number of detections, for each line, is also summarized in Table~\ref{tab:counts_tot_confr}. %The differences are large, amounting to orders of magnitude in some cases.
%However, most of the discrepancies originates from a different terminology. In this paper detection of an AGN means that an AGN-excited line has a flux above the survey threshold while in the \citet{Spin12} terminology it is sufficient that the galaxy containing an AGN is detected, even though the AGN contribution is below threshold.
%These discrepancies, substantially larger than those found in \citet{Bonato2014} for the starburst component, result from the combination of different line to continuum luminosity ratios and different evolutionary models.

The different results reflect the different approaches adopted by the authors (only very marginally the different evolutionary models used). In particular, \citet{Spin12}\footnote{In the calculation of the IR luminosities used to calibrate the line to continuum relations in the \citet{Spin12} paper a factor of 1.8 was inadvartently omitted, resulting in an overestimate of the line counts (an Erratum on that has been submitted to ApJ).} based their estimate on empirical relations between the line luminosities and the total IR luminosity for the Seyfert galaxies of the $\,12\,\mu$m galaxy sample \citep{Tomm08,Tomm10}, and for the pure starburst galaxies of the sample of \citet{Bern09}. These luminosity ratios were then used to convert the total IR luminosity functions from different evolutionary models (i.e. \citealt{Valiante09}; \citealt{Franc10}; \citealt{Grupp11}) into line luminosity functions and to derive the numbers of
objects detectable by SPICA at different redshifts and luminosities. This approach - purely empirical - could be taken as an upper limit for the SPICA detections, since it applies the relations found for a sample of AGNs to all the AGN populations considered in the models, regardless of the AGN contribution to $L_{\rm IR}$, while - thanks to {\em Herschel} - we now know that $L_{\rm IR}$ in IR detected AGNs is typically composed by both AGN and starburst contributions \citep{Hatziminaoglou10} in different proportions
(in large fractions of objects the AGN contributes to $L_{\rm IR}$ for $<$10\%; see \citealt{Delvecchio14}).

On the other hand the approach adopted in this paper requires bolometric corrections, endowed with a substantial uncertanty, to derive the relationships between line and bolometric luminosities. 
%\textbf{For example, if we decrease by a factor of 2 the ratio between the AGN bolometric luminosity and the measured luminosity at 12$\,\mu$m,
%the number of AGNs detectable
%by the SPICA/SAFARI survey of $5\,\hbox{deg}^2$, with 1\,hr integration/FoV, in the 3 typical AGN lines, [NeV]$14.32\mu$m, [NeV]$24.31\mu$m and [OIV]$25.89\mu$m,
%is 19, 143 and 1621, respectively. If, on the other hand, we increase that ratio by the same factor of 2 the detections in the 3 lines drop to 1, 18 and 328, respectively. We note, however, that the effective AGN bolometric correction is strongly constrained by both direct measurements of AGN SEDs and by multi-frequency AGN counts and luminosity functions at different redshifts \citep[see, e.g.,][]{Cai13} so that a factor of 2 is probably a generous upper limit to the uncertainty.}
%Therefore, models and local template SEDs provide the only way to estimate the AGN intrinsic luminosity related to the line emission.
We are now working on a semi-empirical approach, based on SED decomposition (see \citealt{Berta13}; \citealt{Delvecchio14}) of local \citep{Tomm08,Tomm10} and high-z samples of AGN
(from {\it Herschel} PACS Evolutionary Probe: \citealt{Lutz11}), considering nuclear observational data in the MIR for estimating the $L_{\rm line}$/$L_{\rm bol}$ relations
and the AGN accretion luminosity function of \citet{Delvecchio14} to derive expected detection numbers.
The results of that work will be published in a forthcoming paper (Gruppioni, Berta, Spinoglio et al. in preparation).

%%%%%%%%%%%%
% FIGURE %
%%%%%%%%%%%%
%
%%%%%%%%%%%%
% FIGURE %
%%%%%%%%%%%%
\begin{table*} %C13 model - Number of {\bf starforming galaxies} per spectral line for {\bf SPICA/SAFARI}
\centering
\footnotesize
\resizebox{\textwidth}{!}{%
\begin{tabular}{lccccccccc}
\hline
\hline
Spectral line & $0.00-0.75$ & $0.75-1.25$ & $1.25-1.75$ & $1.75-2.25$ & $2.25-2.75$ & $2.75-4.00$ & $4.00-6.00$ & $6.00-8.00$ & All z \\
%\hline
%\multicolumn{9}{l}{\bf Starforming galaxies} \\
\hline
${\rm PAH 6.2}\mu$m  &    0 (0) &    0 (0) &    0 (0) &    0 (0) &    0 (0) &    0 (0) &    0 (432) &    0 (19) &    0 (451)\\
${\rm PAH 7.7}\mu$m  &    0 (0) &    0 (0) &    0 (0) &    0 (0) &    0 (0) &    0 (8750) &    0 (6787) &    0 (237) &    0 (15775)\\
${\rm PAH 8.6}\mu$m  &    0 (0) &    0 (0) &    0 (0) &    0 (0) &    0 (0) &    0 (5778) &    0 (1280) &    0 (26) &    0 (7085)\\
${\rm H_{2}}9.66\mu$m  & 0 (0) &    0 (0) &    0 (0) &    0 (0) &    $0.2$ (18) &   $0.2$ (21) &    0 ($0.7$) &    0 (0) &    $0.4$ (39)\\
${\rm [SIV]}10.49\mu$m  & 0 (0) &    0 (0) &    0 (0) &    0 (9) &    $0.1$ (169) &    0 (119) &    0 (12) &    0 ($0.2$) &    $0.1$ (308)\\
${\rm H_{2}}12.28\mu$m  & 0 (0) &    0 (0) &    0 (0) &    $0.3$ (58) &    0 (26) &    0 (11) &    0 (1) &    0 (0) &    $0.3$ (95)\\
${\rm [NeII]}12.81\mu$m  & 0 (0) &    0 (0) &    1  (490) &    1  (1515) &    $0.4$ (748) &    $0.3$ (495) &    0 (43) &    0 ($0.3$) &    3 (3293)\\
${\rm [NeV]}14.32\mu$m  &  0 (0) &    0 (0) &    9 (9) &    2 (2) &    $0.4$ ($0.4$) &    $0.2$ ($0.2$) &    0 (0) &    0 (0) &    11 (11)\\
${\rm [NeIII]}15.55\mu$m  &  0 (0) &    1 (82) &    10 (493) &    2 (220) &    $0.3$ (92) &    $0.3$ (52) &    0 (1) &    0 (0) &    13 (941)\\
${\rm H_{2}}17.03\mu$m  & 0 (0) &    9 (84) &    10 (100) &    1 (45) &    $0.4$ (20) &    $0.2$ (10) &    0 ($0.3$) &    0 (0) &    20 (260)\\
${\rm [SIII]}18.71\mu$m  &  0 (0) &  13 (2040) &    10 (1125) &    1 (499) &    $0.4$ (261) &    $0.1$ (143) &    0 (10) &    0 ($0.3$) &    25 (4078)\\
${\rm [NeV]}24.31\mu$m  &    28 (28) &    19 (19) &    8 (8) &    1 (1) &    $0.3$ ($0.3$) &    $0.2$ ($0.2$) &    0 (0) &    0 (0) &    55 (55)\\
${\rm [OIV]}25.89\mu$m  &    426 (426) &    227 (227) &    89 (89) &    20 (20) &    1 (1) &    $0.3$ ($0.3$) &    0 (0) &    0 (0) &    762 (762)\\
${\rm [SIII]}33.48\mu$m  &    235 (9876) &    10 (6760) &    1 (2947) &    1 (1289) &    $0.2$ (771) &    $0.1$ (438) &    0 (21) &    0 (0) &    248 (22104)\\
\hline
\hline
\end{tabular}}
\caption{Predicted redshift distributions of AGNs and, in parenthesis, of galaxies as a whole (starburst plus AGN components) detectable by a SPICA/SAFARI survey covering an area of $5\,\hbox{deg}^2$ in 1\,h integration per FoV.}
\label{tab:counts_spica1}
\end{table*}

\begin{table*}
\centering
\footnotesize
\begin{tabular}{lcccccccc}
\hline
\hline
\small{Spectral line} &     \textbf{This work}                             & \multicolumn{3}{c}{Spinoglio et al. (2012)}  \\
                      & \textbf{\small{\citet{Cai13}}} & \small{\citet{Franc10}} & \small{\citet{Grupp11}} & \small{\citet{Valiante09}} \\
%\hline
%\multicolumn{9}{l}{\bf Starforming galaxies} \\
\hline
${\rm [NeII]}12.81\mu$m  & \boldmath$0.3$             & 15                 & 14                 & 42  \\
${\rm [NeV]}14.32\mu$m   & \textbf{1}                 & $0.4$              & 1                  & --  \\
${\rm [NeIII]}15.55\mu$m & \textbf{1}                 & 41                 & 25                 & 179  \\
${\rm H_{2}}17.03\mu$m   & \textbf{2}                 & 0                  & $0.4$              & 6  \\
${\rm [SIII]}18.71\mu$m  & \textbf{3}                 & 3                  & 2                  & 6  \\
${\rm [NeV]}24.31\mu$m   & \textbf{5}                 & 53                 & 143                & -- \\
${\rm [OIV]}25.89\mu$m   & \textbf{76}                & 326                & 623                & 938  \\
${\rm [SIII]}33.48\mu$m  & \textbf{25}                & 244                & 400                & 1103  \\
\hline
\hline
\end{tabular}
\caption{Numbers of $5\,\sigma$ AGN detections by SPICA/SAFARI for a survey covering $0.5\,\hbox{deg}^2$ in 1\,hr integration/FoV, as predicted by the present model (second column, in boldface) compared with those given (only for 8 lines) by \citet{Spin12} for the 3 models used by them. }
\label{tab:counts_tot_confr}
\end{table*}

%%%%%%%%%%%%
% FIGURE %
%%%%%%%%%%%%
%

%%%%%%%%%%%%
% FIGURE %
%%%%%%%%%%%%
\begin{figure*}
%\hspace{+0.0cm}
\makebox[\textwidth][c]{
\includegraphics[trim=2.5cm 0.4cm 3.0cm 0.6cm,clip=true,width=0.48\textwidth, angle=0]{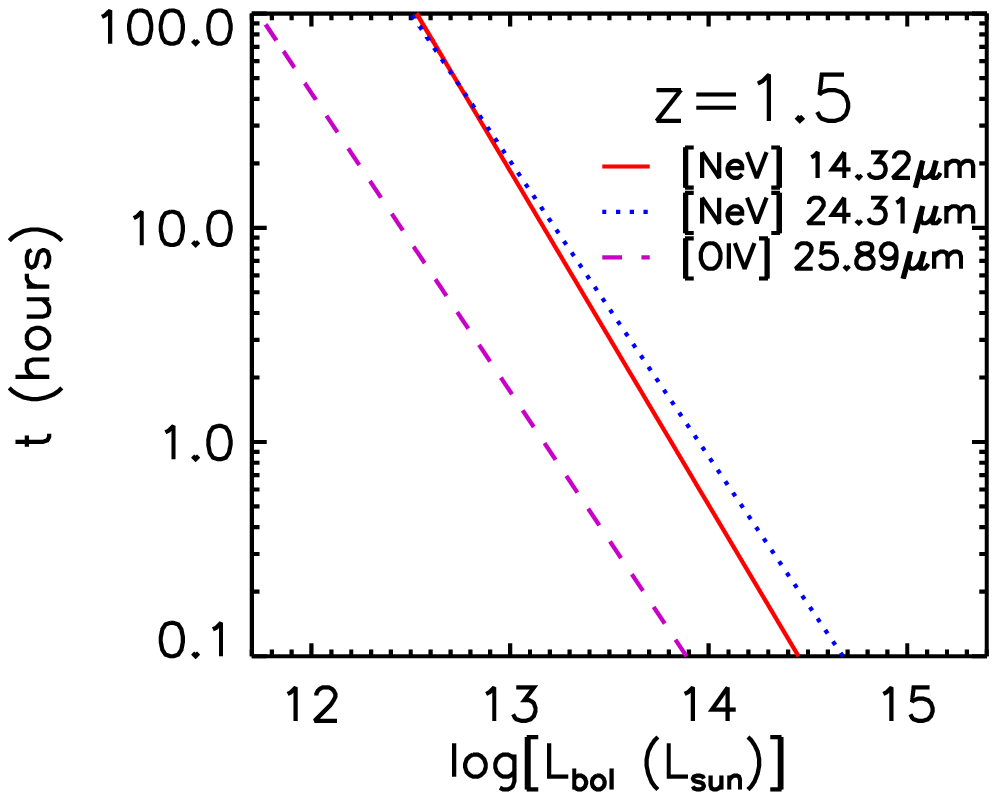}
\includegraphics[trim=2.5cm 0.4cm 3.0cm 0.6cm,clip=true,width=0.48\textwidth, angle=0]{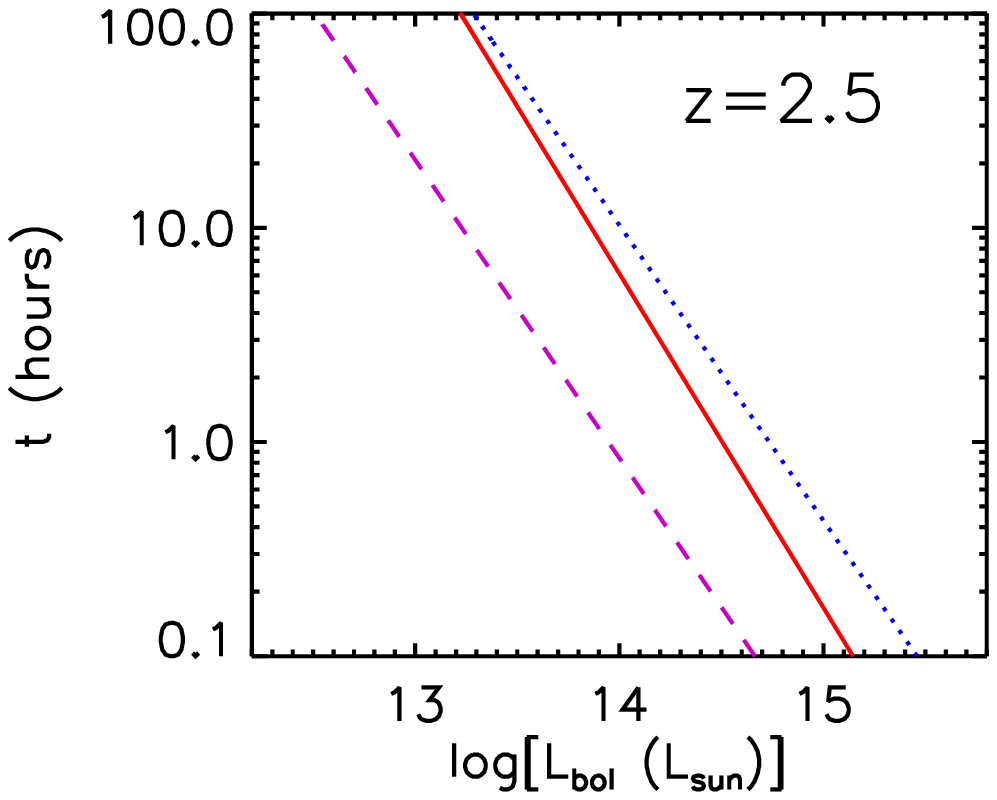}
}
%\vspace{-4.5cm}
\caption{SPICA/SAFARI exposure time per FoV required for a $5\sigma$ line detection of the three typical AGN lines as a function of the AGN bolometric luminosity for $z=1.5$ (left) and $z=2.5$ (right).}
 \label{fig:strategy}
\end{figure*}
%

%%%%%%%%%%%%%%%%%%
\section{Observation strategy}\label{sect:survey}
%%%%%%%%%%%%%%%%%%

Quantitative predictions on the number of detections of AGNs and of galaxies as a whole for a survey of $0.5\,\hbox{deg}^2$ and different integration times per FoV are given in Table~\ref{tab:counts_different_surveys}. As illustrated by Fig.~\ref{fig:intcounts_spica_all}, the integral counts have a slope flatter than 2 at and below the detection limit for 1\,hr integration/FoV. This means that the number of detections at fixed observing time increases more extending the survey area than going deeper. This is particularly true for the 3 typical AGN lines (cf. Fig.~\ref{fig:intcounts_spica}), due to the fact that the AGN contribution to the bolometric luminosity is higher at higher luminosities.
For a survey of 0.5\,deg$^{2}$ and 1\,h integration/FoV the detection rate of AGNs is quite low and the predicted redshift distributions sink down at $z\gsim1$ (cf. Fig.~\ref{fig:redshiftdistr_spica}). To investigate with sufficient statistics the galaxy--AGN co-evolution a deeper or a wider area survey is necessary. For example, a survey of 1\,hr/FoV over 5\,deg$^{2}$ would yield about 760 AGN detections in the [OIV]25.89$\mu$m line, with a total observing time of 4500 hours. A deeper survey with the same observing time (exposure time of 10\,hr/FoV over 0.5\,deg$^{2}$) would give 427 AGN detections (see Table~\ref{tab:counts_different_surveys}).

%So blind surveys have a low efficiency in investigating the AGN activity and the galaxy-AGN co-evolution, especially at $z>1$.
The blind spectroscopic survey may be complemented by follow-up observations of bright high-$z$ galaxies already discovered at (sub-)mm wavelengths over much larger areas. Fig.~\ref{fig:strategy} shows, as an example, the SPICA/SAFARI exposure time per FoV necessary for a $5\sigma$ detection of the three typical AGN lines at $z=1.5$ and $z=2.5$ as a function of the AGN bolometric luminosity. When SPT and \textit{Herschel} survey data will be fully available we will have samples of many hundreds of galaxies with either intrinsic or apparent \citep[i.e. boosted by strong gravitational lensing;][]{Neg10,Neg14,Vieira13} IR luminosities larger than $10^{13}\,L_{\odot}$. These IR luminosities correspond to $\hbox{SFR}> 10^3\,M_\odot\,\hbox{yr}^{-1}$ \citep{KennicuttEvans2012}. Comparing the stellar with the halo mass function \citet{Lapi11} estimated that the active star formation phase lasts for $\sim 0.7\,$Gyr. Thus objects that bright are the likely progenitors of spheroidal galaxies with stellar masses $M_\star \gsim 7\times 10^{11}/\mu\,M_\odot$, $\mu$ being the gravitational magnification factor. {As shown by Fig.~\ref{fig:strategy}, SPICA/SAFARI can detect, in 10\,h, the ${\rm [OIV]}25.89\mu$m line from an AGN with (real or apparent) bolometric luminosity of $\simeq 3\times 10^{12}\,L_\odot$ at $z=1.5$ and of $\simeq 2\times 10^{13}\,L_\odot$ at $z=2.5$ . For Eddington limited accretion these luminosities correspond to black hole masses of $\simeq 9\times 10^7/\mu\,M_\odot$ and $\simeq 6\times 10^8/\mu\,M_\odot$.} For comparison, from the black hole/stellar mass correlation \citep{KormendyHo2013} we get that the final black hole masses associated to spheroidal galaxies with stellar masses of $\simeq 10^{11}\,M_\odot$ and $\simeq 10^{12}\,M_\odot$ are $\simeq 5\times 10^8\,M_\odot$ and $\simeq 7\times 10^9\,M_\odot$, respectively. This means that pointed SPICA/SAFARI observations can allow us to investigate early phases of the galaxy/AGN co-evolution, when the black hole mass was one or even two orders of magnitude lower than the final one.

%The statistics of objects at $z\ge 3$ detected by the SPICA/SAFARI reference survey is very poor. So, in order to investigate the AGN activity at high-$z$, it is expedient to complement the blind spectroscopic survey with follow-up observations of the high-$z$ galaxies already discovered at (sub-)mm wavelengths over much larger areas. Figure~\ref{fig:strategy} shows, as an example, the SPICA/SAFARI exposure time per FoV necessary for a $5\sigma$ detection of the three typical AGN lines detectable at $z=3$ and $z=4$ as a function of the bolometric luminosity. When SPT and \textit{Herschel} survey data will be fully available we will have samples of many hundreds of bright strongly lensed galaxies \citep[apparent luminosities $L_{\rm IR}>10^{13}\,L_{\odot}$;][]{Neg10,Vieira13}. The AGN emission of a source with a $L_{\rm bol, AGN}$ of $10^{13.5}\,L_{\odot}$, for example, can be detected in little more than 1 hour in one line ([OIV]$25.89\mu$m) and in little more than 10 hours in the three AGN lines at $z=3$, while at $z=4$ it needs about 4 hours for the detections in the [OIV]$25.89\mu$m line and about 40 hours in the three lines.

\begin{table} %C13 model - Number of {\bf starforming galaxies} per spectral line for {\bf SPICA/SAFARI}
\centering
\footnotesize
\begin{tabular}{lccc}
\hline
\hline
Spectral line & $t = 1h$ & $t=10\,\hbox{h}$ \\
  & $0.5\,\hbox{deg}^2$ & $0.5\,\hbox{deg}^2$ \\
\hline
${\rm PAH 6.2}\mu$m &  0 (45) & 0 (280) \\
${\rm PAH 7.7}\mu$m &  0 (1578) & 0 (4183) \\
${\rm PAH 8.6}\mu$m &  0 (709) & 0 (2661) \\
${\rm H_{2}}9.66\mu$m & $4\times10^{-2}$ (4) & $0.1$ (33) \\
${\rm [SIV]}10.49\mu$m &  $1\times10^{-2}$ (31) & $3\times10^{-2}$ (156) \\
${\rm H_{2}}12.28\mu$m & $3\times10^{-2}$ (10) & $0.1$ (74) \\
${\rm [NeII]}12.81\mu$m & $0.3$ (330) & 3 (1490) \\
${\rm [NeV]}14.32\mu$m & 1 (1) & 5 (5) \\
${\rm [NeIII]}15.55\mu$m & 1 (94) & 11 (627) \\
${\rm H_{2}}17.03\mu$m & 2 (26) & 9 (236) \\
${\rm [SIII]}18.71\mu$m & 3 (409) & 18 (1979) \\
${\rm [NeV]}24.31\mu$m & 5 (5) & 48 (48) \\
${\rm [OIV]}25.89\mu$m & 76 (76) & 427 (427) \\
${\rm [SIII]}33.48\mu$m &  25 (2211) & 147 (7826) \\
\hline
\hline
\end{tabular}
\caption{AGN and total (in parenthesis) detections obtainable by SPICA-SAFARI surveys covering $0.5\,\hbox{deg}^2$ in 1\,h integration/FoV ($2^{nd}$ column) and 10\,h integration/FoV ($3^{th}$ column).}
\label{tab:counts_different_surveys}
\end{table}
%\end{landscape}

%%%%%%%%%%%%%%%%%%
%\section{Diagnostic diagrams}\label{sect:diagnostic}
%%%%%%%%%%%%%%%%%%

%Figure~\ref{fig:diagnostic1} shows, as an example, the simulated diagnostic diagram $L_{[NeV]14.32\mu\rm m}/L_{[NeII]12.81\mu\rm m}$ versus $L_{[OIV]25.89\mu\rm m}/L_{[NeII]12.81\mu\rm m}$ for the unlensed protospheroidal galaxies, varying the IR AGN emission fraction.

%%%%%%%%%%%%
% FIGURE %
%%%%%%%%%%%%
%\begin{figure*}
%\hspace{+2.0cm}
%\makebox[\textwidth][c]{
%\includegraphics[width=1.55\textwidth, angle=0]{../LFIR/Grafici_articolo/panel_redshdistr.ps}
%\includegraphics[trim=3.3cm 1.0cm 0.7cm 1.3cm,clip=true,width=1.03\textwidth, angle=0]{spheroids_ratio_agn_fraction.eps}
%}
%\vspace{-2.5cm}
%\caption{Diagnostic diagram including 3 lines of our sample, for the unlensed protospheroidal population, varying the AGN contribution to the total IR continuum luminosity of the sources.}
% \label{fig:diagnostic1}
%\end{figure*}
%

%%%%%%%%%%%%%%%%%%
\section{Conclusions}\label{sect:concl}
%%%%%%%%%%%%%%%%%%

%Our understanding of the cosmological evolution of IR galaxies has dramatically improved in recent years. Observational determinations of their IR luminosity functions up to $z\simeq 4$ are now available. A major step forward will be the characterization of their physical properties, such as the intensity of their interstellar radiation fields, their chemical abundances, the temperatures and densities of their ISM. This will be made possible by the IR spectroscopy provided by forthcoming facilities and in particular by the SAFARI instrument on SPICA. To optimize the survey strategy and to understand the redshift and galaxy luminosity ranges that can be measured it is necessary to have detailed predictions of the line luminosity functions as a function of redshift and of the corresponding number counts as a function of line flux.

%The distributions of line luminosities have then been exploited to estimate the redshift dependent line luminosity functions starting from the IR continuum luminosity functions. Simulations have been made to take fully into account the effect of the dispersions in the line to continuum luminosity ratios.

We have improved over earlier estimates of redshift-dependent luminosity functions of IR lines detectable by SPICA/SAFARI by building a model that deals in a self consistent way with emission of galaxies as a whole, including both the starburst and the AGN component. For proto-spheroidal galaxies, that dominate the cosmic star formation rate at $z\gsim 1.5$, we have derived analytic formulae giving the probability that an object at redshift $z$ has a starburst luminosity $L_{\ast}$ or an AGN luminosity $L_{\bullet}$ given the total luminosity of $L_{\rm tot}=L_{\ast}+L_{\bullet}$. Each proto-spheroid of luminosity $L_{\rm tot}$ was assigned IR luminosities of the starburst and of the AGN component based on the above probability distributions. The association of the AGN component to late-type galaxies was made on the basis of the observed correlation between SFR and accretion rate.

The relationships between line and IR luminosities of the starburst component derived by \citet{Bonato2014} have been updated whenever new data have become available in the meantime and additional lines have been taken into account. New relationships between line and AGN bolometric luminosities have been derived.

These ingredients were used to work out predictions for the source counts in 11 mid/far-IR emission lines partially or entirely excited by AGN activity, as well as in the $6.2$, $7.7$ and $8.6\mu$m PAH lines, overlooked by \citet{Bonato2014}.
The expected outcome of a survey covering $5\,\hbox{deg}^2$ with a 1\,hr integration/FoV is specifically discussed. Taking into account also the additional 9 lines, with negligible AGN contribution, investigated by \citet{Bonato2014}, i.e. a total of 23 IR lines, we have estimated the number of objects detectable in at least 2, 3, 4, 5 and 6 lines. We estimate that such survey will yield about 760 AGN detections in the [OIV]$25.89\mu$m line. A deeper survey (10\,hr/FoV over 0.5\,deg$^{2}$) requiring the same observing time (4500 hours) will yield about $\sim430$ detections in the same line.

The AGN contribution to the detectability of galaxies as a whole is minor. We show that this is due to the combination of two factors: the rarity of bright AGNs, due to their short lifetime, and the relatively low (compared to starbursts) line to bolometric luminosity ratios.

%This conclusion looks at odds with the estimates by \citet{Spin12} who, based on other evolutionary models, claim a much higher number of AGN detections.
%, but with a different meaning. While in this paper detection of an AGN means that an AGN-excited line has a flux above the survey threshold, according to \citet{Spin12} it is sufficient that the galaxy containing an AGN is detected, even though the AGN contribution is below threshold.

We also recommend pointed observations of the brightest (either because are endowed with extreme SFRs or because luminosities are strongly magnified by gravitational lensing) galaxies previously detected by large area surveys such as those by \textit{Herschel} and by the SPT. Such observations will allow us to investigate early phases of the galaxy/AGN co-evolution when the black hole mass was much lower than the final one, given by the present day correlation with the stellar mass.

\section*{Acknowledgements} We are grateful to the anonymous referee for many constructive comments that helped us improving this paper and to Prof. Kotaro Kohno and Prof. Takehiko Wada for pointing out the importance of the $6.2$, $7.7$ and $8.6\mu$m PAH lines, overlooked by our previous study. We acknowledge financial support from ASI/INAF Agreement 2014-024-R.0 for the Planck LFI Activity of Phase E2 and from PRIN INAF 2012, project ``Looking into the dust-obscured phase of galaxy formation through cosmic zoom lenses in the Herschel Astrophysical Large Area Survey''.

%%%%%%%%%%%%%
% BIBLIOGRAPHY %
%%%%%%%%%%%%%

\end{document}